\documentclass[aps,prd,twocolumn,showpacs,10pt,superscriptaddress,preprintnumbers]{revtex4-1}

\usepackage{amsmath}
\usepackage{graphicx}
\usepackage{amssymb}
\usepackage{color} 

\usepackage{ulem}
\usepackage{textcomp} 
\usepackage{slashed}
\usepackage{epsfig,latexsym,cancel,amssymb,amsmath,verbatim,mathrsfs}
\usepackage{color}
\usepackage{graphicx}
\usepackage{enumitem}

\newcommand{\beq}{\begin{equation}}
\newcommand{\eeq}{\end{equation}}
\newcommand{\bea}{\begin{eqnarray}}
\newcommand{\eea}{\end{eqnarray}}
\newcommand{\nn}{\nonumber}
\newcommand{\GeV}{\mathrm{GeV}}

\usepackage{ulem}

\begin{document}
\preprint{
	{\vbox {
			\hbox{\bf LA-UR-21-22425}
}}}

\title{Loop induced top quark FCNC through top quark and dark matter interactions}

\author{Yandong Liu}
\email{ydliu@bnu.edu.cn}
\affiliation{Key Laboratory of Beam Technology of Ministry of Education, College of Nuclear Science and Technology, Beijing Normal University, Beijing 100875, China}
\affiliation{Beijing Radiation Center, Beijing 100875, China}

\author{Bin Yan}
\email{binyan@lanl.gov}
\affiliation{Theoretical Division, Group T-2, MS B283, Los Alamos National Laboratory, P.O. Box 1663, Los Alamos, NM 87545, USA}

\author{Rui Zhang}
\email{zhangr@ihep.ac.cn}
\affiliation{Theoretical Physics Division, Institute of High Energy Physics, Beijing 100049, China}

\begin{abstract}
We present a comprehensive analysis of the loop induced top quark  FCNC signals at the LHC within one class of the simplified model. The loop level FCNC interactions are well motivated to avoid the hierarchy of the top quark couplings from the new physics and standard model. Such a theory will posit a Majorana dark matter candidate and could be tested through dark matter relic density, direct detection experiments  (the scattering between dark matter and  heavy nuclei), and  the collider signals at the LHC. 
We find that the spin-independent (SI) scattering between Majorana dark matter and nuclei will vanish at the leading order, while the next-to-leading order correction to the SI scattering becomes significant to constrain the parameter space of the model.  A detailed comparison between direct detection experiments and LHC searches is also discussed and both of them are very important to fully constrain the model.
\end{abstract}

\maketitle

\section{Introduction}
Top quark flavor changing neutral current (FCNC)  processes can only occur via loops in the standard model (SM) and their production rates are highly suppressed by the Glashow-Iliopoulos-Maiani mechanism~\cite{Glashow:1970gm}. For example, the branching ratios of  top quark decay $t\to Xq$, where $X=h, Z,\gamma, g$ and  $q=u,c$ are around $10^{-16}\sim 10^{-12}$ in the SM~\cite{AguilarSaavedra:2004wm}. As a result,  it would be a challenge to observe the SM top quark FCNC effects with the current or foreseeable dataset. 
However, the testable top quark FCNC signals could be induced in the new physics (NP) beyond the SM at the tree or loop level, e.g. ${\rm BR}(t\to Xq)\sim \mathcal{O}(10^{-5})$ in the minimal supersymmetry standard model (MSSM)~\cite{AguilarSaavedra:2004wm} or two Higgs doublet model~\cite{Gaitan:2015hga,Abbas:2015cua,Arhrib:2015maa}. Therefore, any evidence of the top quark FCNC signals would shed light on the various possible NP models and also the underlying mechanism of generating top quark FCNC. 

The top quark FCNC effects have been extensively studied by the theoretical~\cite{Oyulmaz:2018irs,Shi:2019epw,Khanpour:2019qnw,Cao:2007bx,Yang:2008sb,Zhang:2008yn,Han:2009zm,Yue:2011an,Gao:2011fx,Cao:2014udj,Sun:2014qoa,Liu:2015kkp,Sun:2016kek,Wang:2017pdg} and experimental~\cite{Khachatryan:2015att,Aad:2019pxo,Abe:1997fz,Khachatryan:2016sib,Aad:2015gea,Aaltonen:2008qr,Abazov:2010qk,Chatrchyan:2013nwa,Sirunyan:2017kkr,Aaboud:2018nyl,Aad:2015uza,Aaltonen:2008ac,Abazov:2011qf,Khachatryan:2016atv,Sirunyan:2017uae,Aaboud:2018oqm} groups within the SM effective field theory (SMEFT) framework or specific NP models. Both the ATLAS and CMS collaborations show that the current dataset could put a strong constraint on the ${\rm BR}(t\to Xq)$ through top quark rare decay and single top quark production. For example, the most stringent limits on the branching ratios of  ${\rm BR}(t\to \gamma u)<6.1\times 10^{-5}$ and ${\rm BR}(t\to \gamma c)<2.2\times 10^{-4}$ at the 95\% confidence level (CL) are set by the single top quark  and a photon production~\cite{Aad:2019pxo} at the 13 TeV LHC. The limit of the ${\rm BR}(t\to g q)$ is comparable to the  ${\rm BR}(t\to \gamma q)$ and it shows that the upper limit is $2\times 10^{-5}$ and $2\times 10^{-4}$ for $u$ and $c$ quark, respectively at the LHC Run-I~\cite{Khachatryan:2016sib,Aad:2015gea}.  Both the top quark decay~\cite{Khachatryan:2016sib,Aad:2015gea,Khachatryan:2016atv,Aaboud:2018oqm} and single top quark associated with a gauge boson production~\cite{Sirunyan:2017kkr,Sirunyan:2017uae} could constrain the branching ratios ${\rm BR}(t\to Z/hq)$ and the most stringent limits come from the top quark decay, i.e.  ${\rm BR}(t\to Zu)<1.7\times 10^{-4}$, ${\rm BR}(t\to Zc)<2.4\times 10^{-4}$, ${\rm BR}(t\to hu)<1.1\times 10^{-3}$ and ${\rm BR}(t\to hc)<1.2\times 10^{-3}$. 

The SMEFT  is a powerful framework to parametrize potential NP effects and widely applied in the top quark FCNC phenomenology~\cite{Lavoura:2003xp,Degrande:2014tta,Durieux:2014xla}.  
The corresponding Wilson coefficients are constrained to be around $\mathcal{O}(10^{-3})\sim \mathcal{O}(10^{-2})$ with the NP scale $\Lambda=1~{\rm TeV}$. With higher luminosity data being accumulated,  an order of magnitude improvement on the top quark FCNC anomalous couplings is expected at the high-luminosity LHC (HL-LHC), operating at the $\sqrt{s}=14~{\rm TeV}$ with an integrated luminosity of $3~{\rm ab}^{-1}$. Thus it is timely to check those NP models which could induce the top quark FCNC anomalous couplings. 

The top quark FCNC anomalous couplings could be generated through the tree or loop level in NP models. It is well know that the multiplet Higgs models could generate tree level top quark FCNC couplings if the Yukawa structure violates the Glashow-Weinberg-Paschos (GWP) theorem~\cite{Glashow:1976nt,Paschos:1976ay},  e.g. the general two Higgs double models. 
However, such models have been constrained seriously by top quark FCNC data, and as a result{\color{yellow},} a large hierarchy was generated between top quark FCNC and other top quark couplings (e.g. $tbW$~\cite{Cao:2015doa,Cao:2015qta} and $Zt\bar{t}$~\cite{Cao:2020npb}). This situation could be relaxed or avoided if we require the top quark FCNC couplings to be generated only in the loop level, which is similar to the case of the loop induced neutrino mass models~\cite{Ma:1998dn,Cao:2017xgk}.  For example,  the top quark FCNC couplings could be generated by triangle loop diagram through the squarks mixing effects in the MSSM and its value agrees well with the LHC limitation as the mass scale of $\mathcal{O}(1)~{\rm TeV}$ for the heavy particles in the loop, without unduly small  NP couplings.

One obvious way of generating loop induced top quark FCNC couplings is introducing an additional $\mathcal{Z}_2$ symmetry to avoid possible tree level interactions, i.e.  new heavy particles are odd under the $\mathcal{Z}_2$, while the SM particles are even.  One byproduct of such models is that the lightest $\mathcal{Z}_2$ odd neutral particle could be a promising dark matter (DM) candidate. Therefore, it is useful to combine dark matter direct detection and the collider searches  to constrain or search this kind of the NP models. Comparing to the UV complete theories such as MSSM, we would like to use a simplified model to show the idea in this work. The simplified models include a finite number of parameters with a reasonable simple theoretical framework and have became a robust mechanism to search various NP particles. In this work, we will focus on the fermionic dark matter and scalar mediators.  It is also possible to introduce a dark gauge boson in the loop, however, which is beyond the scope of this work.  We show that the simplified model could generate the sizable top quark FCNC anomalous couplings and their predictions are consistent with the present LHC data and could be tested at the HL-LHC and future hadron colliders. In addition, the parameter space is  also testable  by the future dark matter direct detection experiments, (e.g. XENON20T~\cite{Aprile:2020vtw} and DARWIN~\cite{Aalbers:2016jon}) and dark matter searches at the LHC.

\section{Model Setup}
In this section,  we discuss the  detail of our simplified model. It contains a fermionic dark matter ($\chi$) and scalar mediator particles which interact with the dark matter and the SM quarks. The dark matter  is a singlet under the SM gauge groups and can be either a Dirac or a Majorana fermion and we will focus on the  Majorana case in this work.  In general, the top quark from the NP interactions could be both left and right-handed under the $SU(2)_L$ gauge symmetry. However, the left-handed top and bottom quarks transform as doublet under $SU(2)_L$, and as a result, any modification to the left-handed top quark couplings will also induce corrections to $B$ physics, e.g.~\cite{Blanke:2017tnb,Blanke:2017fum}. Therefore, we will focus on the right-handed top quark case in order to avoid the possible constraints from precise measurements of $B$ physics. The gauge representations of the mediators under the SM gauge group $SU(3)_C\otimes SU(2)_L\otimes U(1)_Y$ are,
\beq
\Phi_1\sim (3,1,2/3),\quad \Phi_2\sim (3,2,1/6).
\eeq
The general gauge invariant Lagrangian can be written as following
\begin{align}
&\mathcal{L}_{\rm eff} \supset    (\mathrm{D}_\mu \Phi_1)^\dagger (\mathrm{D}^\mu \Phi_1) +
 (\mathrm{D}_\mu \Phi_2)^\dagger (\mathrm{D}^\mu \Phi_2) - M^2_2 \Phi_2^\dagger \Phi_2  \nonumber \\
 &- M^2_1 \Phi_1^\dagger \Phi_1 + (Y_{RI} \bar{\chi}P_R U_I \Phi_1^\ast+ Y_{LI} \bar{Q}_{I} P_R \chi \Phi_2+ \mbox{h.c.}) \nonumber \\
 &+\frac{1}{2} \bar{\chi}(\not{\!{\partial}} -m_\chi) \chi  - (\mu   \Phi_1^\ast H \Phi_2 + \mbox{h.c.})+..., \label{case:I}
\end{align}
where $U_I$ and $Q_I$ are the right- and left-handed quarks under $SU(2)_L$ with the family index $I=1,2,3$, respectively. $H$ is the Higgs doublet field and defined as $H^T=\frac{1}{\sqrt{2}}(0,v+h)$ with $v=246~{\rm GeV}$ in the unitarity gauge. The  covariant derivative in Eq.~\ref{case:I} is defined as,
\beq
D_\mu=\partial_\mu-ig_s G_\mu^a T^a-ig W_\mu^i\tau^i-ig^\prime Y B_\mu.
\eeq
Here $G_\mu^a$, $W_\mu^i$ and $B_\mu$ are the $SU(3)_c$, $SU(2)_L$ and $U(1)_Y$ gauge fields, respectively, and $g_s$, $g$ and $g^\prime$ are their gauge couplings. The gauge parameters $g$ and $g^\prime$ are related to the electric charge and the Weinberg mixing angle by $gs_W=g^\prime c_W=e$, where $s_W=\sin\theta_W$ and $c_W=\cos\theta_W$. Furthermore, $T^a$ and $\tau^i$ are the $SU(3)_c$ and $SU(2)_L$ generators and $Y$ is the hypercharge, in the representation of the field on which the derivative acts. 

After the electroweak symmetry breaking, we should introduce quark transform matrices ($V^{R,L}$) in the Yukawa sector of the new mediators due to the misalignment of the weak and mass eigenstate of the quarks in the SM, i.e.
\begin{align}
\mathcal{L}_{\rm Yukawa}=Y_{RI} V^{R}_{Ii} \bar{\chi}P_R u_i \Phi_1^\ast+ Y_{LI} V^L_{Ii} \bar{q}_{i} P_R \chi \Phi_2+ \mbox{h.c.}
\end{align}
To avoid the possible limits from the low energy data, the maximal flavor violating assumption~\cite{Bar-Shalom:2007xeu,Bernreuther:2008ju}  will be used in the analysis.
Such ansatz could be realized through some flavor symmetry assumption~\cite{Linster:2018avp} or some turning of the underlying model parameters. Here, we will not restrict ourselves to any underlying model, and instead take a phenomenological approach towards the maxima flavor violation case.   The effective Yukawa couplings which are related to the top quark FCNC anomalous couplings could be parametrized as,
\begin{align}
\mathcal{L}_{\rm Yukawa}=y_t  \bar{\chi}P_R t \Phi_1^\ast+ y_{u}  \bar{q}_{u} P_R \chi \Phi_2+ \mbox{h.c.},
\end{align}
where $t$ and $q_u^T=(u,d)$ are the right-handed top quark and left-handed first generation quarks, respectively. It is straightforward to generalize our results to the second generation light quarks.  A similar simplified model with other assumptions are also widely discussed in the literature and could be found  in Refs.~\cite{DiFranzo:2013vra,Kilic:2015vka,Mohan:2019zrk}.

After the electroweak symmetry breaking, the colored doublet scalar $\Phi_2^T=(\phi_u,\phi_d)$ will mix with $\Phi_1$ through the 3-scalar interaction.
We define the mass eigenstates as $\phi_1, \phi_2$ which can relate to the gauge eigenstates by
\begin{align}
\left(
\begin{array}{c}
\phi_1 \\
\phi_2
\end{array}
\right) 
=\left( \begin{array}{cc} 
\cos \theta & -\sin \theta \\
\sin\theta & \cos\theta
\end{array}
\right)  
 \left( \begin{array}{c} 
\Phi_1 \\
\phi_u
\end{array}
\right).
\end{align}
The mixing angle $\theta$ is defined as
\begin{align} \label{Eq:mixing}
\sin 2\theta = \frac{\sqrt{2}\mu v}{m_2^2-m_1^2},
\end{align}
where $m_{1,2}$ are the mass of the two physics scalars, 
\begin{align}
m_1^2 = \frac{1}{2}(M_1^2+M_2^2-\sqrt{(M_1^2-M_2^2)^2+2\mu^2 v^2}), \label{EQ:m1} \\
m_2^2=\frac{1}{2}(M_1^2+M_2^2+\sqrt{(M_1^2-M_2^2)^2+2\mu^2 v^2}). \label{EQ:m2}
\end{align}
The mass of scalar $\phi_d$ is same as $M_2$ and can be related to the $m_{1,2}$ by
\begin{align}
M_2^2 = m_2^2\cos^2\theta + m_1^2 \sin^2\theta. 
\end{align}

\section{Loop induced top quark FCNC}
\label{sec:FCNC}
\begin{figure}
\centering
\includegraphics[scale=0.5]{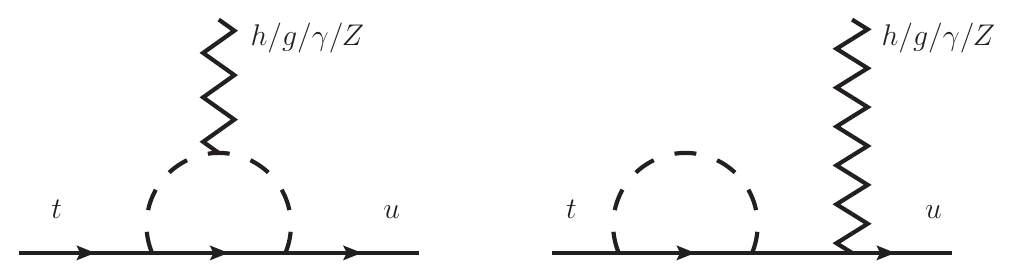}
\caption{\label{Fig:TopFCNC}Illustrative one-loop Feynman diagrams to generate top quark FCNC anomalous couplings. The left diagram depicts the vertex correction and the right one is corresponding to the quark field mixing.}
\end{figure}
The top quark FCNC anomalous couplings could be generated through the triangle and self-energy diagrams in our simplified model; see Fig.~\ref{Fig:TopFCNC}. And the interactions can be parameterized by the effective operators under the $SU(3)_C\otimes U(1)_{EM}${\color{yellow}:}
\begin{align}
\mathcal{L}_{\rm eff}&=\lambda_{tqh}\bar{q}_L t_R h+\frac{\lambda_{tq\gamma}}{m_t}\bar{q}_L\sigma_{\mu\nu}t_R A^{\mu\nu}+\frac{\lambda_{tqg}}{m_t}\bar{q}_L\sigma_{\mu\nu}t_R G^{a\mu\nu}T^a\nn\\
&+\frac{\lambda_{tqZ}^{(1)}}{m_t}\bar{q}_L\sigma_{\mu\nu}t_R Z^{\mu\nu}+\lambda_{tqZ}^{(2)}\bar{q}_L\gamma_\mu t_L Z^{\mu}+h.c.{\color{yellow}.}
\end{align}
Note that the gauge invariance of $U(1)_{EM}$ forbids the vector and axial-vector current interactions of  $tq\gamma$ and $tqg$. The matching coefficients could be obtained from one-loop calculations and the results are,
\begin{widetext}
\begin{align}
\lambda_{tqh}&=\frac{y_t y_u m_\chi}{16\sqrt{2}\pi^2} \left\lbrace  \frac{\sin 2\theta}{\sqrt{2}v} \left(B_{\chi 1}^0 -B_{\chi 2}^0 \right)  +
 \mu \left( \sin^4 \theta - \sin^2 \theta \cos^2 \theta \right) C_{\chi 1 2}^{0ht} +\right.\nn\\
&\left. \mu \left( \cos^4 \theta - \sin^2 \theta \cos^2 \theta \right) C_{\chi 2 1}^{0ht} + 
 2\mu \sin^2 \theta \cos^2 \theta C_{\chi 1 1}^{0ht}+
 2\mu  \sin^2 \theta \cos^2 \theta C_{\chi 2 2}^{0ht}
   \right\rbrace, \label{EQ:tqh} \\
 \lambda_{tq\gamma} &= \frac{Qe}{32\pi^2} \frac{m_\chi}{m_t} y_ty_u \sin\theta\cos\theta( B_{\chi 2}^t-B_{\chi 2}^0-B_{\chi 1}^t+B_{\chi 1}^0),\\
 \lambda_{tqg} &= \frac{g_s}{32\pi^2} \frac{m_\chi}{m_t} y_ty_u \sin\theta\cos\theta( B_{\chi 2}^t-B_{\chi 2}^0-B_{\chi 1}^t+B_{\chi 1}^0),\\
 \lambda_{tqZ}^{(1)} &=\frac{g}{16\pi^2c_W}\frac{m_\chi m_t}{4(m_t^2-m_Z^2)}y_ty_u \sin2\theta\left\lbrace (B_{\chi 1}^0-B_{\chi 2}^0)f_R-\frac{m_t^2+m_Z^2}{m_t^2-m_Z^2}f_L(B_{\chi 1}^t-B_{\chi 2}^t)\right.\nn\\
 &\left.+\frac{m_Z^2}{m_t^2-m_Z^2}\left[(2f_L-\cos^2\theta)B_{11}^Z-(2f_L-\sin^2\theta)B_{22}^Z+\cos2\theta B_{12}^Z\right]\right.\nn\\
 &\left.+\frac{m_Z^2}{2(m_t^2-m_Z^2)}\left[(2f_L-\cos^2\theta)\lambda(m_1)C_{1\chi 1}^{0tZ}-(2f_L-\sin^2\theta)\lambda(m_2)C_{2\chi 2}^{0tZ}\right]\right.\nn\\
 &\left.+\frac{\cos^2\theta}{4(m_t^2-m_Z^2)}\left[2(m_1^2-m_2^2)m_t^2+(\lambda(m_1)+\lambda(m_2))m_Z^2\right]C_{2\chi 1}^{0tZ}\right.\nn\\
 &\left.+\frac{\sin^2\theta}{4(m_t^2-m_Z^2)}\left[2(m_1^2-m_2^2)m_t^2-(\lambda(m_1)+\lambda(m_2))m_Z^2\right]C_{1\chi 2}^{0tZ}
 \right\rbrace \\
 \lambda_{tqZ}^{(2)} &=-\frac{g}{16\pi^2 c_W}\frac{m_\chi}{m_t}y_ty_uf_L\sin\theta \cos\theta\left(B_{\chi 2}^t-B_{\chi 2}^0-B_{\chi 1}^t+B_{\chi 1}^0\right)+2\lambda_{tqZ}^{(1)}. \label{EQ:tqZ}
\end{align}
\end{widetext}
where
\begin{align}
f_L&=\frac{1}{2}-\frac{2}{3}s_W^2,&
f_R&=-\frac{2}{3}s_W^2,&
\lambda(m_i)&=2m_\chi^2-2m_i^2-m_t^2+m_Z^2.
\end{align} 
We abbreviate the scalar functions as $B_{ij}^a=B_0(m_a^2,m_i^2,m_j^2)$ and $C_{ijk}^{abc}=C_0(m_a^2,m_b^2,m_c^2,m_i^2,m_j^2,m_k^2)$, and their definitions can be found in Ref.~\cite{tHooft:1978jhc}. Note that $m_{a,b,c}=0$ for $a,b,c=0$.

The matching coefficients could be much simplified in the decouple limt, i.e. $m_{1,2,\chi}\sim \Lambda\gg m_{t,h,Z}$,
\begin{align}
\lambda_{tqh}&\simeq-\frac{1}{384\sqrt{2}\pi^2} y_t y_u \mu   \frac{m_t^2}{\Lambda^3}, \label{EQ:tqh}\\
\lambda_{tq\gamma} &\simeq\frac{iQ e}{392\sqrt{2}\pi^2} y_t y_u \frac{\mu v m_t}{\Lambda^3},\\
\lambda_{tqg} &\simeq\frac{ig_s}{392\sqrt{2}\pi^2} y_t y_u \frac{\mu v m_t}{\Lambda^3}, \label{Eq:tqg} \\
\lambda_{tqZ}^{(1)} &\simeq-\frac{g y_t y_u  (f_L+f_R)}{768\sqrt{2}\pi^2c_W}\frac{\mu v m_t}{\Lambda^3},\\
\lambda_{tqZ}^{(2)} &\simeq\frac{g y_t y_u  (f_L-f_R)}{384\sqrt{2}\pi^2c_W}\frac{\mu v m_t}{\Lambda^3}.
\label{eq:lam}
\end{align}
We should note that the scale $\mu\sim v$ in above limit, see Eq.~\ref{Eq:mixing}.  

It is also worth noting that the coefficient $\lambda_{tqh}$ could be enhanced by the triple scalar interaction in a  large $\mu$ limit (see Eq.~\ref{EQ:tqh}), which could be generated by a large mass splitting between $\phi_1$ and $\phi_2$ for a fixed mixing angle $\theta$.  Therefore, it is interesting to consider the limit of $m_2 \gg m_{1, \chi}\sim m_{t, h, Z}$ with a fixed mixing angle $\theta$. It shows  that the coefficients could be
\begin{align}
\lambda_{tqh} &\simeq -\frac{13y_uy_t m_2^2\sin^32\theta}{1536\pi^2 m_\chi v}\,, \label{eq:tqh3}\\
\lambda_{tq\gamma}& \simeq -\frac{y_u y_t Qe (2-\pi/\sqrt{3}) \sin 2\theta}{32\pi^2}\,, \label{eq:tqa3}\\
\lambda_{tqg}& \simeq -\frac{y_u y_t g_s (2-\pi/\sqrt{3}) \sin 2\theta}{32\pi^2}\,, \label{eq:tqg3}.
\end{align} 
It is evident that $\lambda_{tqh}$ will be enhanced by the mass $m_2$, while the other coefficients ($\lambda_{tq\gamma/g/Z}$) approach a constant for a large $m_2$. The result for $\lambda_{tqZ}$ is still complicated in this limit and will not be shown in here.

\begin{figure*}
\centering
\includegraphics[scale=0.57]{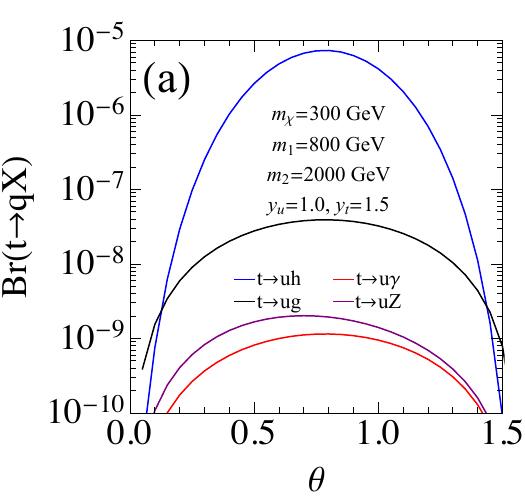}
\includegraphics[scale=0.58]{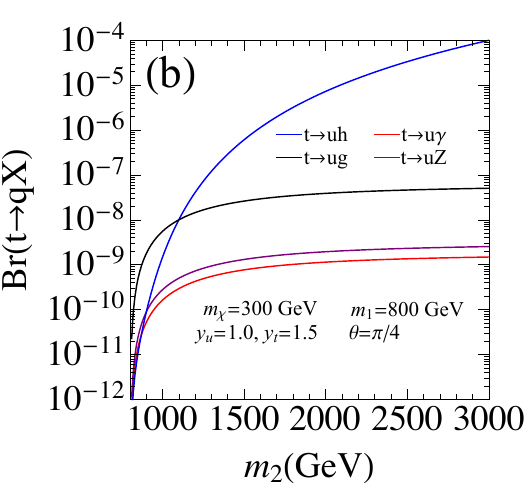}
\includegraphics[scale=0.55]{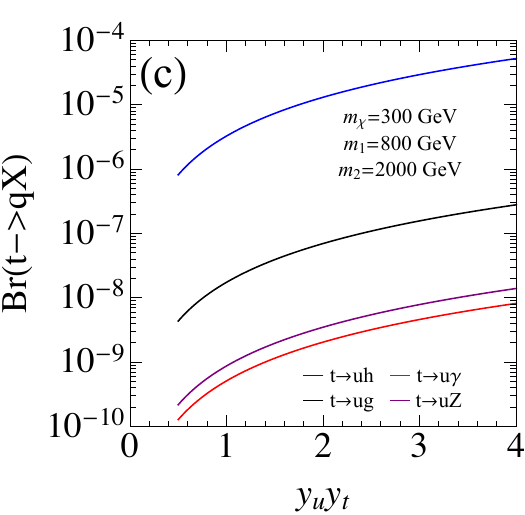}
\caption{\label{Fig:topdecaybr}The top quark FCNC decay branching ratios in our simplified model.}
\end{figure*}

The partial decay widths of top quark through the FCNC anomalous couplings are
\begin{align}
\Gamma(t\to qh)&=\lambda_{tqh}^2\frac{m_t}{32\pi}\left(1-\frac{m_h^2}{m_t^2}\right)^2,\\
\Gamma(t\to q\gamma) &=\lambda_{tq\gamma}^2\frac{m_t}{4\pi},\\
\Gamma(t\to qg) &=\lambda_{tqg}^2\frac{m_t}{4\pi}4C_F,\\
\Gamma(t\to qZ)&=\frac{m_t}{32\pi}\left(1-\frac{m_Z^2}{m_t^2}\right)^2\left[4\left(\lambda_{tqZ}^{(1)}\right)^2\left(2+\frac{m_Z^2}{m_t^2}\right)\right.\nn\\
&\left.-12\lambda_{tqZ}^{(1)}\lambda_{tqZ}^{(2)}+\left(\lambda_{tqZ}^{(2)}\right)^2\left(2+\frac{m_t^2}{m_Z^2}\right)\right].
\label{eq:gamma}
\end{align}
Figure~\ref{Fig:topdecaybr} shows the decay branching ratios from the top quark FCNC anomalous couplings  with the benchmark parameters, namely $m_\chi=300~\GeV$, $\theta=\pi/4$, $y_u=1.0$ and $y_t=1.5$. 
The Wilson coefficients $\lambda_{tqh/\gamma/g/Z}$ can be found in Eqs.~\ref{EQ:tqh}-\ref{EQ:tqZ} and are calculated by LoopTools~\cite{Hahn:1998yk}.
Figure~\ref{Fig:topdecaybr}(a) displays the top quark FCNC decay branching ratios as a function of mixing angle $\theta$ with scalar mass of $m_1=800~\GeV$ and $m_2=2000~\GeV$.  It is evident that the branching ratios will reach the maximal when the mixing angle $\theta\sim \pi/4$. 
This is because those top quark FCNC couplings (see Eqs.~\ref{EQ:tqh}-\ref{EQ:tqZ}) could be generated only through the mixing between $\phi_1$ and $\phi_2$, and,  as a result, the maximal mixing of $\theta=\pi/4$ will induce the largest FCNC couplings. 
The typical branching ratios of $t\to q+Z/\gamma/g$ are around $\mathcal{O}(10^{-9})-\mathcal{O}(10^{-8})$. However, the ${\rm BR}(t\to qh)$ could be reached around  $\mathcal{O}(10^{-5})$ due to the $\lambda_{\rm tqh}$ will be enhanced by $m_2^2$ in the limit of $m_2 \gg m_1$; see Eq.~\ref{eq:tqh3}. Figure~\ref{Fig:topdecaybr}(b) depicts the top quark FCNC decay branching ratios as a function of $m_2$. 
It is evident that ${\rm BR}(t\to qh)$ is strongly dependent on the $m_2$, but  ${\rm BR}(t\to q\gamma/g/Z)$ are not in  a large $m_2$ region, as argued before; see Eqs.~\ref{eq:tqh3}-\ref{eq:tqg3}. In the limit of $m_2\gg m_1$, the ${\rm BR}(t\to qh)$ could be around  $\mathcal{O}(10^{-5})$, which is large enough and  can be detected at the HL-LHC and future hadron colliders.   However, it would be a challenge to probe other top quark FCNC signals of this model due to the small decay branch ratios.

\begin{figure*}
\centering
\includegraphics[scale=0.295]{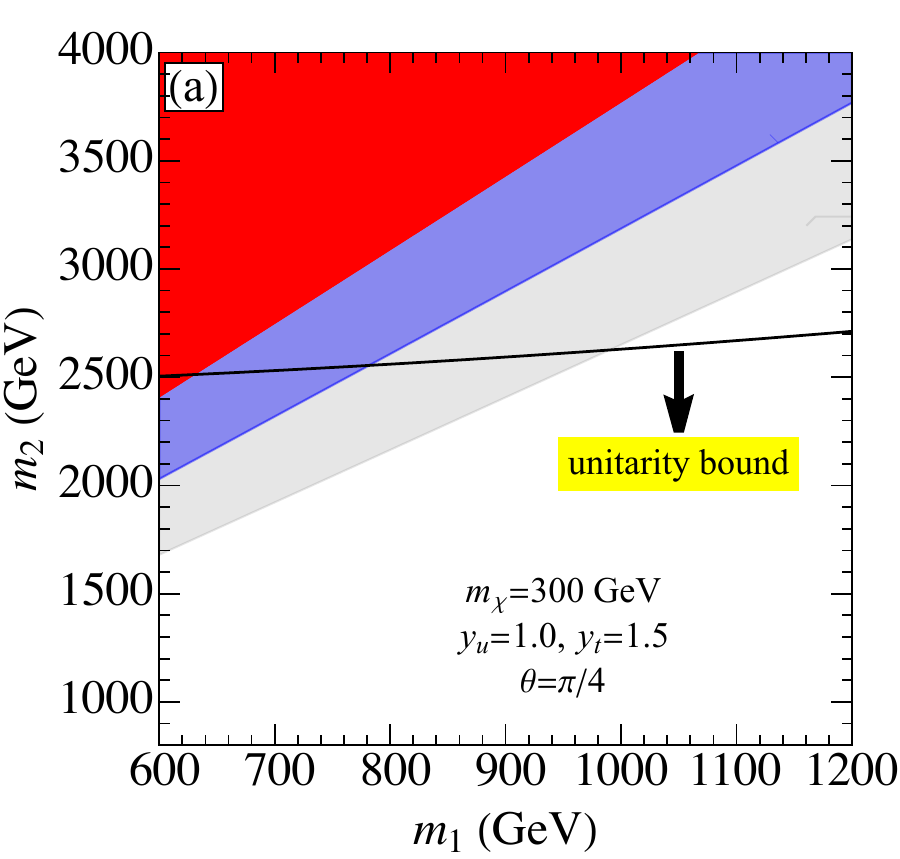}
\includegraphics[scale=0.28]{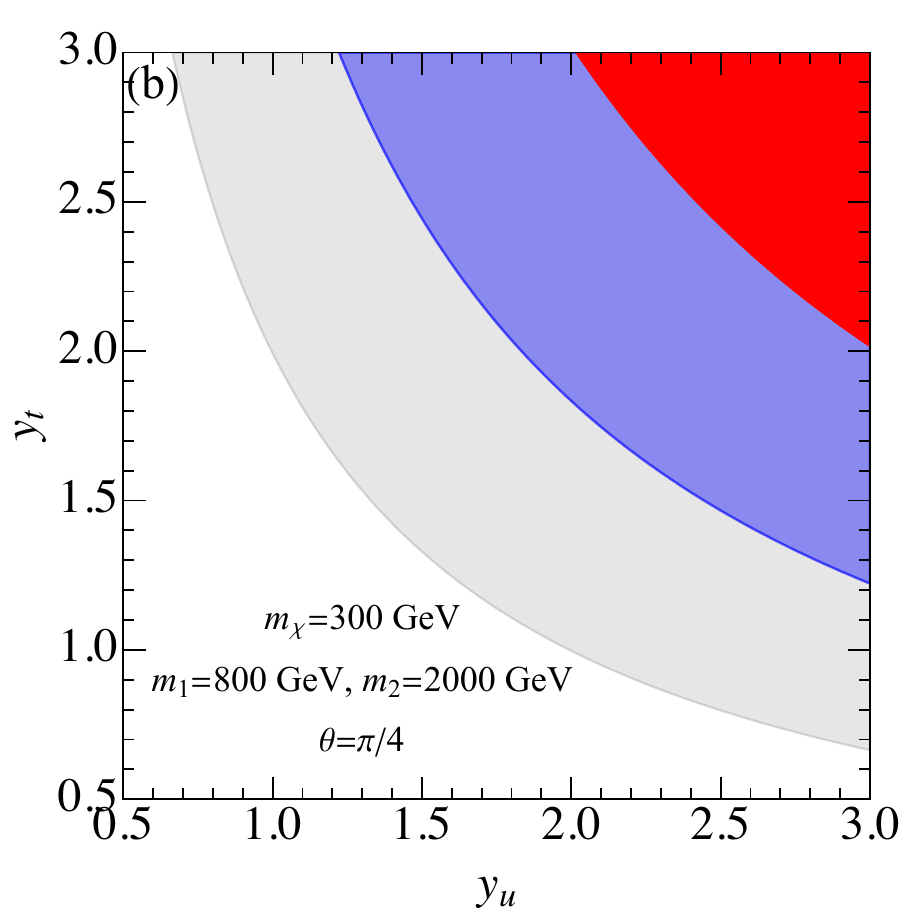}
\includegraphics[scale=0.29]{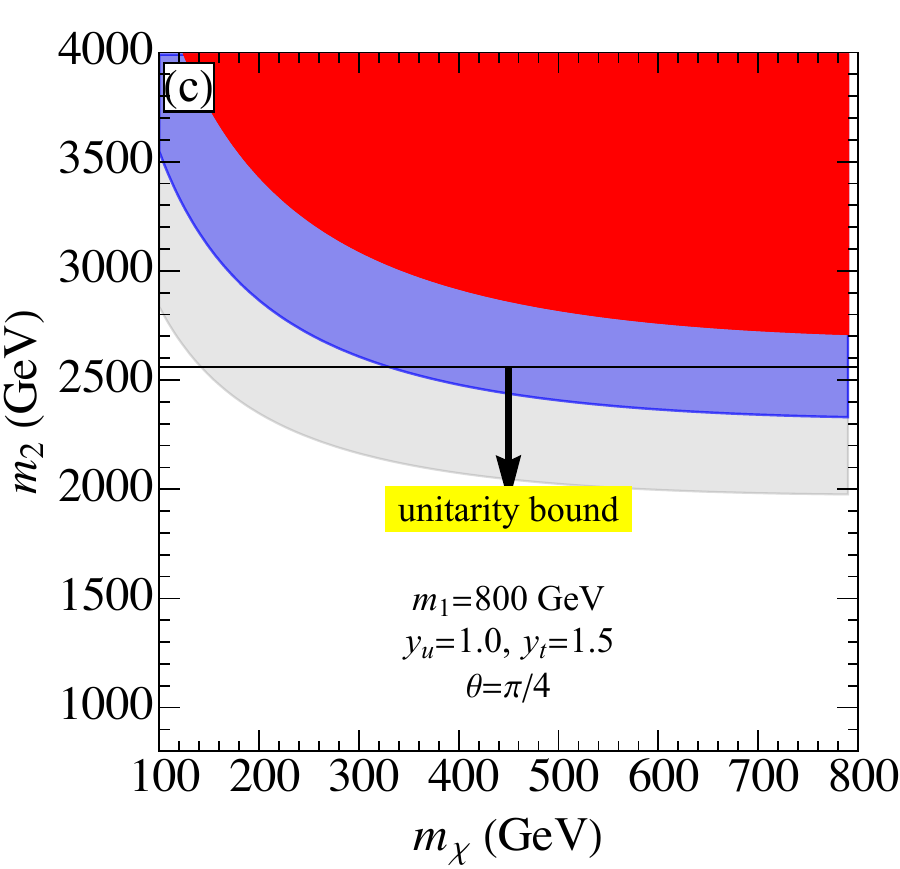}
\includegraphics[scale=0.29]{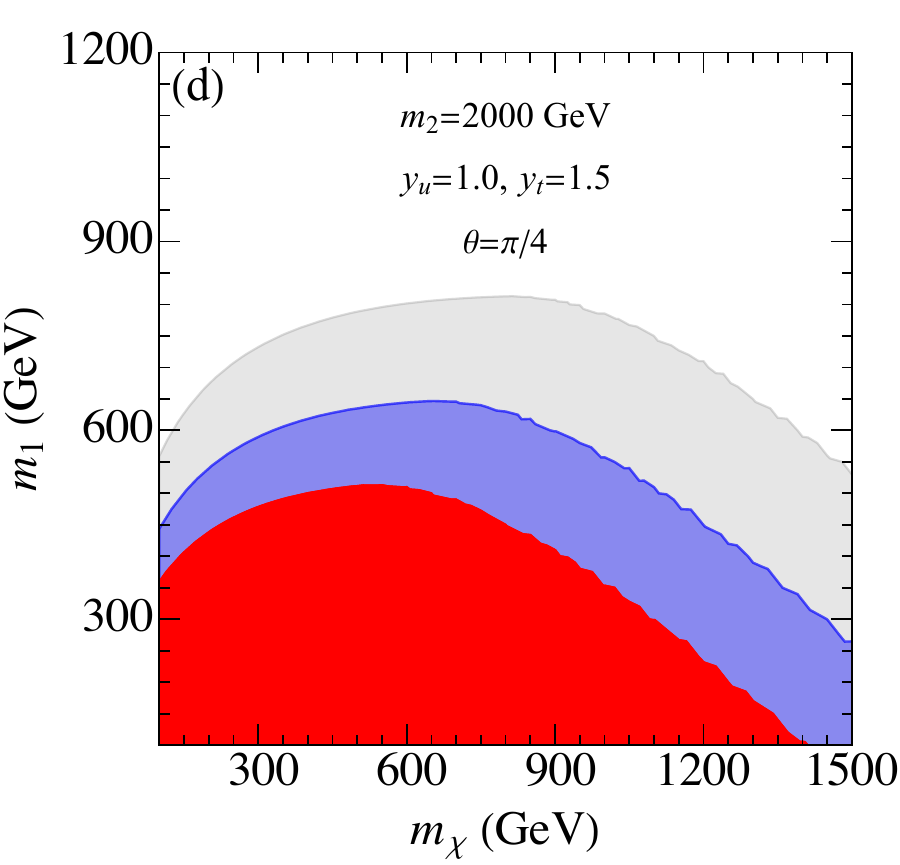}
\caption{\label{Fig:tqh}The expected sensitivity to the model parameters through $t\to qh$ from the HL-LHC (red), HE-LHC (blue) and FCC-HH (gray) at 95\% C.L.. The black line denotes the unitarity bound from the scattering $\phi_i \phi_j^\ast \rightarrow W^+_LW^-_L$.}
\end{figure*}

Figure~\ref{Fig:tqh} displays the testable parameter space in the plane of the model parameters at the 95\% C.L. from $tqh$ measurement at the HL-LHC (red)~\cite{TheATLAScollaboration:2013nbo,ATLAS:2016qxw}, HE-LHC (blue)~\cite{Zhang:2020naz} and FCC-HH (gray)~\cite{Zhang:2020naz}. From Fig.~\ref{Fig:tqh}(a), it is clear that a larger $m_2$ will enhance the loop contribution to $t\to qh$. The contour behavior of Fig.~\ref{Fig:tqh}(b) arises from the fact that the $\lambda_{tqh}\propto y_uy_t$. We also note that a small $m_\chi$ will weaken the loop contribution to $t\to qh$, while a moderate $m_\chi$ will result in a larger ${\rm BR}(t\to qh)$; see Fig.~\ref{Fig:tqh}(c).  

We notice that the scale $\mu$ will be constrained by the unitarity of the scattering $\phi_i \phi_j^\ast \rightarrow W^+_LW^-_L$ ($i,j=1, 2$), with longitudinal polarized $W$ boson. Based on the analysis of the partial wave expansion~\cite{Chanowitz:1978mv}, for the $s$-wave,
\begin{align}
a=\frac{1}{64\pi^2}\int d\Omega \mathcal{M}(\phi_i\phi_j^\ast \rightarrow W_L W_L),
\end{align}
and  the coupled matrix is 
\begin{align}
a=\begin{pmatrix}
0& \frac{\sqrt{2}\mu\delta_{m,n}}{32 \pi v} \\
-\frac{\sqrt{2}\mu\delta_{m,n}}{32 \pi v} &0
\end{pmatrix},
\end{align}
where $\delta_{m,n}$ is the color index. The unitarity of the scattering amplitude requires that the absolute value of the eigenvalues from above coupled matrix should be small than 1~\cite{Lee:1977yc,Lee:1977eg,Chanowitz:1978mv,Chanowitz:1978uj}. Therefore, the upper limit of $\mu$ is
\begin{align}
|\mu| \lesssim  17 ~\rm{TeV}.
\end{align}
We can also translate the upper limit of $\mu$ to the mass splitting between $\phi_1$ and $\phi_2$ through Eq.~\ref{Eq:mixing}; see the black line in Fig.~\ref{Fig:tqh}.  We should note that the unitarity bound would be stronger than the limit from HL-LHC in some parameter space. 

\section{Dark Matter}
In this section we discuss the constraints from the DM relic abundance and the direct detection experiments on our simplified model. 
\subsection{relic density}
The DM relic abundance has been measured by Planck with a high accuracy, i.e. $\Omega_\chi h^2 = 0.1200 \pm 0.0012$ \cite{Aghanim:2018eyx}.  It requires that the theoretical predicted relic abundance from the simplified model should be smaller than the observed value to avoid dark matter overproduction. Assuming DM particle is thermally produced in the early universe,  $\Omega_\chi h^2$ is totally determined by the thermally averaged annihilation cross sections $\langle \sigma v_{\rm rel} \rangle$. The evaluation of DM density is controlled by the Boltzmann equation and its relic abundance is given by
\begin{align}
\Omega_\chi h^2 \approx 2.76\times 10^{8} Y \frac{m_\chi}{\rm GeV},
\end{align}
where $Y$ is the DM number density of today in the comoving frame, 
\begin{align}
 Y = \sqrt{\frac{45}{\pi}} \frac{g_{\ast}(x_f)^{1/2}/g_{\ast s}(x_f)}{m_{\rm PL} m_\chi \langle \sigma v_{\rm rel} \rangle} x_f.
 \end{align}
 Here $x_f\equiv m_\chi/T_f$, with $T_f$ denoting the decoupling temperature and $m_{\rm PL}$ is the Planck mass. $g_\ast(x_f)$ and $g_{\ast s}(x_f)$ is the effective energy and entropy degree of freedom at the temperature of DM decoupling, respectively.  
 Therefore, the minimal of the $\langle \sigma v_{\rm rel} \rangle$ should be
\begin{align}
\langle \sigma v_{\rm rel} \rangle &\approx 0.71   \frac{0.12}{\Omega_\chi h^2}  \frac{x_f}{25}\frac{g_{\ast}(x_f)^{1/2}/g_{\ast s}(x_f)}{0.1} ~\rm{pb}.
\end{align}
The annihilation cross section can be expanded based on the velocity, i.e. $\langle \sigma v_{\rm rel} \rangle  \doteq a + b\langle v_{\rm rel}^2 \rangle+\mathcal{O}(\langle v_{\rm rel}^4 \rangle)$. The coefficient $a$ and $b$ describes the contributions from s-wave and p-wave scattering, respectively.  Since  $\langle v_{\rm rel}^2 \rangle \sim 0.3$ during the DM decoupling, the s-wave scattering will play the key role for the DM annihilation.

\begin{figure}
\centering
\includegraphics[scale=0.6]{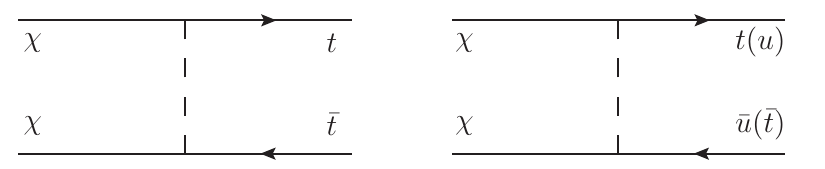}
\caption{\label{DM:Majorana:annihilation} Majorana dark matter annihilation processes.}
\end{figure}

In the simplified model, the dominant contributions to DM annihilation come from 
s-wave processes $\chi\bar{\chi}\to t\bar{t}, t\bar{u}, u\bar{t}$ and p-wave processes $\chi\bar{\chi}\to u\bar{u}, d\bar{d}$;
see Fig.~\ref{DM:Majorana:annihilation}. The s-wave contributions from $u\bar{u}$ and $d\bar{d}$ are suppressed by the quark mass due to the flip of the quark helicity and are ignored in the analysis. The annihilation cross sections are given as following,
\begin{widetext}
\begin{align}
&\langle \sigma v_{\rm rel} \rangle_{t\bar{t}} \simeq  \frac{N_C  y_t^4 m_t^2 (2 m_\chi^2+(m_2^2-m_1^2)\cos(2\theta) +m_1^2+m_2^2 -2 m_t^2 )^2}{128\pi (m_\chi^2 + m_1^2 -m_t^2)^2 (m_\chi^2 + m_2^2 -m_t^2)^2} \sqrt{1-\frac{m_t^2}{m_\chi^2}}, \\
&\langle \sigma v_{\rm rel} \rangle_{t\bar{u}+u\bar{t}} \simeq \frac{ N_C  y_t^2 y_u^2 m_\chi^2 (m_1^2-m_2^2)^2 \sin^2(2\theta)  }{2\pi (2m_\chi^2 + 2m_1^2 -m_t^2)^2 (2m_\chi^2 + 2m_2^2 -m_t^2)^2} (1 - \frac{m_t^2}{4m_\chi^2})^2, \\
&\langle \sigma v_{\rm rel} \rangle_{u\bar{u}} \simeq \frac{y_u^4 }{64 \pi (m_\chi^2+m_1^2)^4 (m_\chi^2+m_2^2)^4} \{ m_\chi^2 m_1^4m_2^4 \Delta^2 +8 m_\chi^4 m_1^4m_2^4 \Delta \nonumber \\
 &~~~~~~~~~~~~+m_\chi^6 [( m_1^4+m_2^4) \Delta^2 +20 m_1^4 m_2^4]  + 4 m_\chi^8 [( m_1^2+m_2^2) \Delta^2 +4 m_1^2 m_2^2 (m_2^2-m_1^2) \cos(2\theta)] \nonumber\\
 &~~~~~~~~~~~~  +m_\chi^{10} [5 \Delta^2 + 4( m_1^4 +m_2^4)] + 8m_\chi^{12}\Delta + 4  m_\chi^{14}
  \}v_{\rm rel}^2,\\
  &\langle \sigma v_{\rm rel} \rangle_{d\bar{d}} \simeq \frac{y_u^4 m_\chi^2 (m_\chi^4 + M_2^4)}{16\pi (m_\chi^2+M_2^2)^4} v_{\rm rel}^2,
\end{align}
\end{widetext}
where $N_C=3$ denotes the color factor of quarks and $\Delta=m_1^2+m_2^2 - (m_2^2-m_1^2)\cos(2\theta)$.  As we expected that the annihilation from $t\bar{t}$ mode contributes to the s-wave and the rate is proportional to $m_t$ due to the flip of the quark helicity, while $u\bar{u}$ and $d\bar{d}$ modes will only contribute to the p-wave when we ignore their mass. Furthermore, the annihilation to $t\bar{u}$ and $u\bar{t}$ final states come from the mixing of the scalars $\Phi_1$ and $\Phi_2$, thus the rates are proportional to the mixing angle $\sin(2\theta)$.
 
\begin{figure*}
\centering
\includegraphics[scale=0.56]{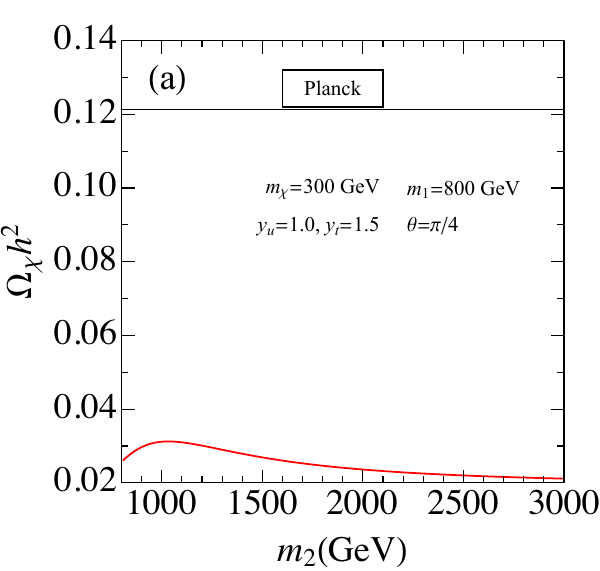}
\includegraphics[scale=0.56]{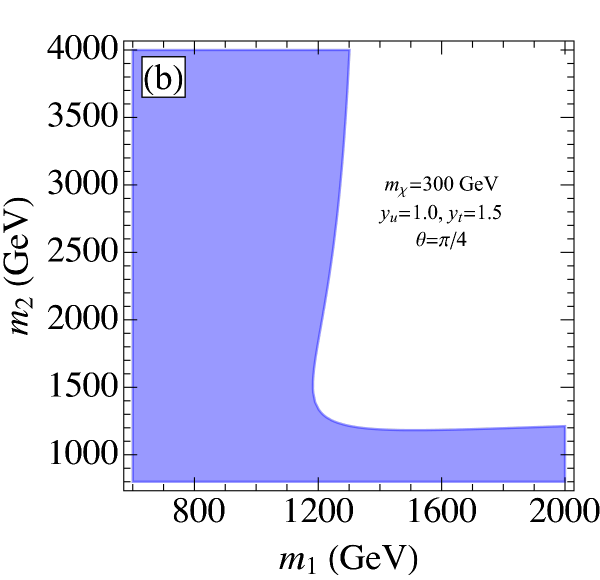}
\includegraphics[scale=0.56]{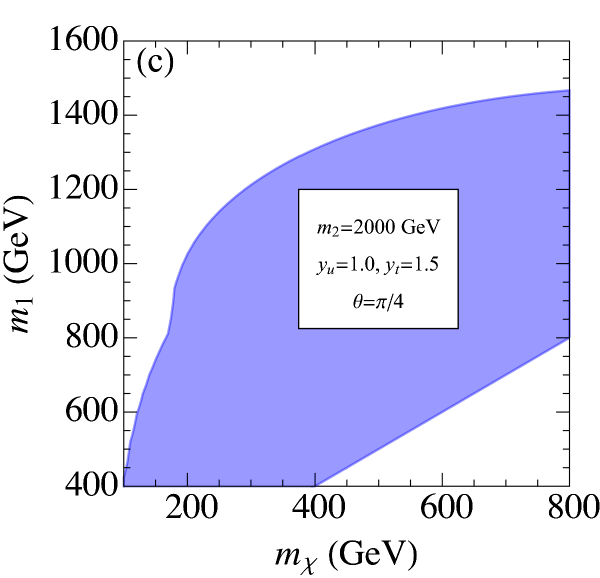}
\caption{\label{Fig:rd}Dark matter relic abundance and limits from Planck 2018~\cite{Aghanim:2018eyx}}
\end{figure*}

In Fig.~\ref{Fig:rd}(a), we show the relic density in our simplified model with the benchmark parameters $m_\chi=300~{\rm GeV},~ m_1=800~{\rm GeV}, y_u=1, y_t=1.5$ and $\theta=\pi/4$. The black line represents the upper limit from Planck 2018 at $2\sigma$ level~\cite{Aghanim:2018eyx}.  We note that the relic density approaches a constant in the large $m_2$  region due to the decoupling behavior of $\phi_2$. Its effects could be understood from the  thermal cross sections $\langle \sigma v_{\rm rel} \rangle_{t\bar{u}+u\bar{t}}$ and $\langle \sigma v_{\rm rel} \rangle_{t\bar{t}}$ in the limit of  $m_2\to\infty${\color{yellow}:}
\begin{align}
&\langle \sigma v_{\rm rel} \rangle_{t\bar{t}} \simeq \frac{N_C y_t^4 m_t^2 (1 + \cos(2\theta))^2}{128\pi (m_1 +m_\chi-m_t)^2} \sqrt{1-m_t^2/m_\chi^2} , \\
&\langle \sigma v_{\rm rel} \rangle_{t\bar{u}+u\bar{t}} \simeq \frac{N_C y_t^2 y_u^2 m_\chi^2 \sin^2(2\theta)}{8\pi (2m_1^2+2m_\chi^2-m_t^2)^2 }\sqrt{1-m_t^2/4m_\chi^2}.
\end{align}
We show the allowed parameter space on the plane $(m_1,m_2)$ and $(m_\chi,m_1)$ from Planck 2018 at $2\sigma$ level~\cite{Aghanim:2018eyx} in Fig.~\ref{Fig:rd}(b) and Fig.~\ref{Fig:rd}(c), respectively. It shows that the parameter space will be constrained seriously by the DM relic density measurement  in the decoupling limit of $m_{1,2}\gg m_\chi$ .

\subsection{direct detection}
\begin{figure}[h]
\centering
\includegraphics[scale=0.5]{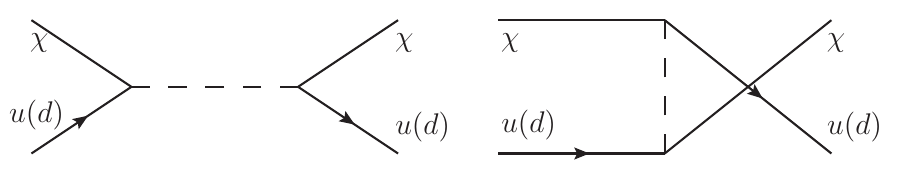}
\caption{\label{DM:dir}DM direct detection process at the tree level.}
\end{figure}

Direct detection is one of cornerstones of DM searches. The goal of direct detection experiments is to detect the rare scattering between the non-relativistic DM and the target material; see Fig.~\ref{DM:dir}.
The scattering cross sections could be separated into spin-independent (SI) and spin-dependent (SD) contributions  based on the interactions between DM and the nucleus.
Since the SI scattering resolves the entire nucleus coherently, the cross section will be enhanced by the squared number of scattering centers (nucleons). While for the SD scattering, DM couples to the nucleon spin and the coherent enhancement effect disappears. To describe the SI and SD scatterings, we use the following effective Lagrangian~\cite{Hisano:2010ct,Hill:2014yka},
\begin{align}
\mathcal{L} =\sum_{u,d} f_q^{\rm{SD}}\bar{\chi} \gamma^\mu \gamma^5 \chi \bar{q} \gamma_\mu \gamma^5 q + \sum_{u,d,s}\mathcal{L}^{\rm{EFT}}_q + \mathcal{L}^{\rm{EFT}}_g,
\end{align}
where
\begin{align}
 \mathcal{L}^{\rm{EFT}}_q &=f_q m_q \bar{\chi}\chi \bar{q}{q}+\frac{g^{(1)}_q}{2m_\chi}\bar{\chi} i(\partial^\mu \gamma^\nu + \partial^\nu \gamma^\mu) \chi O^{(2)}_{q,\mu\nu}\nn\\
&+ \frac{g^{(2)}_q}{m_\chi^2} \bar{\chi}(i\partial^\mu)(i\partial^\nu) \chi O^{(2)}_{q,\mu\nu}, \\
 \mathcal{L}^{\rm{EFT}}_g &= f_G \bar{\chi}\chi G_{~\mu\nu}^{a} G^{a \mu\nu}+\frac{g^{(1)}_G}{2m_\chi}\bar{\chi} i(\partial^\mu \gamma^\nu + \partial^\nu \gamma^\mu) \chi O^{(2)}_{G,\mu\nu}\nn\\ 
&+ \frac{g^{(2)}_G}{m_\chi^2} \bar{\chi}(i\partial^\mu)(i\partial^\nu) \chi O^{(2)}_{G,\mu\nu}.
\end{align}
The twist-2 operators $O^{(2)}_{q,\mu\nu}$ and $O^{(2)}_{G,\mu\nu}$ are defined as
\begin{align}
& O^{(2)}_{q,\mu\nu} \equiv \frac{1}{2} \bar{q} (  \gamma^{ \{\mu \ } iD_{-}^{\ \nu\}} -\frac{g^{\mu\nu}}{4} i \slashed{D}), \\
& O^{(2)}_{G,\mu\nu} \equiv  - G^{a\mu\lambda}G^{a\nu}_{~~\lambda}+\frac{g^{\mu\nu}}{4} (G^{a}_{~\alpha\beta})^2.
\end{align}
Here
\begin{align}
 \gamma^{ \{\mu \ } iD_{-}^{\ \nu\}}=\frac{1}{2}i(\gamma^\mu D_-^\nu+D_-^\mu\gamma^\nu),\quad D_-^\mu=D^\mu-\overleftarrow{D}^\mu.
\end{align}
The Wilson coefficients $f_q^{\rm SD}$, $f_j$ and $g_j^{(i)}$ with $j=q, G$ and $i=1,2$ could match to our simplified model. It is clear that $f_q^{\rm{SD}}$ term 
describes the SD scattering, while $\mathcal{L}^{\rm{EFT}}_q$ and $\mathcal{L}^{\rm{EFT}}_g$ contribute to the SI process. 
The matrix element of DM SD scattering with a nucleon target ($N= p$ or $n$) is
\beq
f^{\rm SD}=\sum_{u,d} f_q^{\rm SD}\Delta_N^q.
\eeq
The hadronic matrix element $\Delta_N^q$ is defined through
\beq
\langle N |\bar{q}\gamma^\mu \gamma^5 q | N \rangle = 2s^\mu \Delta^q_N,
\eeq
where  $s^\mu$ is the spin of the nucleon.  Meanwhile, the matrix element of the SI scattering is
\begin{align}
f_N/m_N =&\sum_{u,d,s}f_{Tq}f_q + \sum_{u,d,s,c,b} \frac{3}{4} [ q(2)+\bar{q}(2)] (g_q^{(1)}+g_q^{(2)}) \nonumber \\
&-\frac{8\pi}{9\alpha_s} f_{TG}f_{G}  + \frac{3}{4} G(2) (g_G^{(1)}+g_G^{(2)}),
\end{align} 
where $m_N$ is the nucleon mass and $f_{Tq}$, $f_{TG}$, $q(2)$, $\bar{q}(2)$ and $G(2)$ are hadronic matrix elements{\color{yellow}:} 
\begin{align}
&\langle N | m_q \bar{q}q | N \rangle  = f_{Tq} m_N, \\
&\langle N |  G_{~\mu\nu}^{a} G^{a \mu\nu}  | N \rangle = - \frac{8\pi}{9\alpha_s} f_{TG}m_N, \\ 
&\langle N | O^{(2)}_{q,\mu\nu}  | N \rangle = \frac{1}{m_N} (p_\mu p_\nu - \frac{1}{4} m_N^2 g_{\mu\nu}) (q(2)+\bar{q}(2)), \\
&\langle N | O^{(2)}_{G ,\mu\nu}  | N \rangle = \frac{1}{m_N} (p_\mu p_\nu - \frac{1}{4} m_N^2 g_{\mu\nu})  G(2).
\end{align}
Here $p$ is the momentum of the nucleon.

With these conventions, the SD and SI cross sections can be written as
\begin{align}
&\sigma_{\chi N}^{\rm{SD}}=\frac{12 \mu_N^2}{\pi} (f^{\rm{SD}})^2,
&\sigma_{\chi N}^{\rm{SI}} = \frac{4 \mu_N^2}{\pi} f_N^2,
\end{align}
where $\mu_N$ is the reduced mass of  the nucleon and DM, i.e. $\mu_N = m_Nm_\chi/(m_N+m_\chi)$.
The SD scattering is dominated by the tree level interactions and the matching coefficient $f_q^{\rm SD}$ could be got at the leading order with the large mass expansion of the mediators $\phi_{1,2,d}$. The detail of the calculation could be found in Refs.~\cite{DiFranzo:2013vra,Mohan:2019zrk} and the results are
\begin{align}
&f^{\rm{SD}}_u =  -\frac{y_u^2}{8}\left[ \frac{\sin^2\theta}{m_\chi^2-m_1^2}+\frac{\cos^2\theta}{m_\chi^2-m_2^2} \right], \\
&f^{\rm{SD}}_d= -\frac{y_u^2}{8}\frac{1}{m_\chi^2-M_2^2}.
\end{align}
While the leading order cross section for the SI scattering will vanish due to the Majorana nature of the DM, as a result, it is necessary to go beyond the leading order in order to estimate the rate of SI scattering. The matching coefficients $f_q$ and $g_q^{(1)}$ could be got from the next-to-leading order expansion of the mediator $\phi_{1,2,d}$ propagators, i.e.
\begin{align}
f_u=& \frac{y_u^2 \sin^2\theta}{16} \frac{m_\chi}{(m_\chi^2-m_1^2)^2} + \frac{y_u^2 \cos^2\theta}{16} \frac{m_\chi}{(m_\chi^2-m_2^2)^2}, \\
f_d =&  \frac{y_u^2 }{16}   \frac{m_\chi}{(m_\chi^2-M_2^2)^2}, \\
g^{(1)}_u = &\frac{y_u^2 \sin^2\theta}{4} \frac{m_\chi}{(m_\chi^2-m_1^2)^2} + \frac{y_u^2 \cos^2\theta}{4} \frac{m_\chi}{(m_\chi^2-m_2^2)^2}, \\
g^{(1)}_d = & \frac{y_u^2 }{4}   \frac{m_\chi}{(m_\chi^2-M_2^2)^2}.
\end{align}
It shows that the SI matching coefficients  $f_q$ and $g_q^{(1)}$ are suppressed by factor $1/(m^2-m_\chi^2)$, with $m=m_{1,2}, M_2$ compared to the $f_q^{\rm SD}$ in the region of we are interested in, i.e. $m_\chi\ll m_{1,2},M_2$.

The SI matching coefficients $f_G$ and $g_G^{(1,2)}$ could be induced at the one loop and can be calculated by the usual Feynman diagrams with the help of the projection operators. Alternatively, one can also use the Fock-Schwinger gauge to simplify the calculation~\cite{Novikov:1983gd,Hisano:2010ct,Mohan:2019zrk}, i.e. $x^\mu A_\mu=0$.  In this work, we will focus on the Fock-Schwinger gauge method and refer the reader to Ref.~\cite{Mohan:2019zrk} for the detail of projection operator approach. In the Fock-Schwinger gauge, one can express the gluon field in terms of its field strength tensor $G_{\mu\nu}$ and maintain explicit gauge invariance for each step in the calculation. The Wilson coefficients can be extracted from the one loop correction to a Majorana fermion propagator through a non-zero background of gluon fields and it shows
\begin{widetext}
\begin{align}
f_{G} =& \frac{\alpha_s  }{32\pi} \Big\{ y_t^2 m_t^2 {\big [}   \int \frac{d^4q}{i \pi^2} \frac{(\slashed{p}+\slashed{q}) \cos^2 \theta }{(q^2-m_1^2)[ (p+q)^2-m_t^2]^4} +     \int \frac{d^4q}{i \pi^2} \frac{(\slashed{p}+\slashed{q}) \sin^2\theta}{(q^2-m_2^2)[ (p+q)^2-m_t^2]^4} {\big]}  \nonumber \\
 & +  y_t^2  m_1^2   \int \frac{d^4q}{i \pi^2} \frac{( \slashed{p}+\slashed{q})\cos^2\theta}{[(q^2-m_1^2)]^4 [(p+q)^2-m_t^2}   + y_t^2  m_2^2  \int \frac{d^4q}{i \pi^2} \frac{(\slashed{p}+\slashed{q})\sin^2\theta}{[(q^2-m_2^2)]^4 [(p+q)^2-m_t^2]}  \nonumber \\
 &+  y_u^2 m_1^2   \int \frac{d^4q}{i \pi^2} \frac{(\slashed{p}+\slashed{q})\sin^2\theta}{[(q^2-m_1^2)]^4 (p+q)^2}  +y_u^2 m_2^2   \int \frac{d^4q}{i \pi^2} \frac{(\slashed{p}+\slashed{q})\cos^2\theta}{[(q^2-m_2^2)]^4 (p+q)^2}  
+y_u^2  M_2^2 \int \frac{d^4q}{i \pi^2} \frac{\slashed{p}+\slashed{q}}{[(q^2-M_2^2)]^4 (p+q)^2}  \Big \}.
\label{eq:fg}
\end{align}
\end{widetext}

The loop integration in Eq.~\ref{eq:fg} can be calculated analytically and we show the detail in the Appendix.

The Wilson coefficients of gluon twist-2 operators could be obtained from the scattering matrix element,
\begin{align}
&M_{\rm twist-2}=2\pi\alpha_s\bar{\chi}[f(y_t\cos\theta,m_t,m_1)+f(y_t\sin\theta,m_t,m_2)\nn\\
&+f(y_u\sin\theta,0,m_1)+f(y_u\cos\theta,0,m_2)+f(y_u,0,M_2)]\chi O^{(2)}_{G,\mu\nu}
\end{align}
where the function $f(g_\chi,m_q,m_\phi)$ is defined as
\begin{align}
f(g_\chi,m_q,m_\phi)&= 
 g_\chi^2\int\frac{d^4q}{(2\pi)^2} \frac{\gamma^\mu(p+q)^\nu +\gamma^\nu(p+q)^\mu }{(q^2-m_\phi^2)[ (p+q)^2-m_q^2]^3}  \nonumber \\
&+g_\chi^2\int\frac{d^4q}{(2\pi)^2} \frac{(\slashed{p}+\slashed{q}) [-m_q^2 g^{\mu\nu} - 2 (p+q)^\mu (p+q)^\nu ]}{(q^2-m_\phi^2)[ (p+q)^2-m_q^2]^4} \nonumber \\
&+\frac{1}{2} g_\chi^2\int\frac{d^4q}{(2\pi)^2} \frac{(\slashed{p}+\slashed{q}) [ (q^2-m_\phi^2)g^{\mu\nu}-4q^\mu q^\nu ]}{ (q^2-m_\phi^2)^4 [(p+q)^2-m_q^2]}.
\end{align}
Therefore, the twist-2 Wilson coefficients are
\begin{align}
g_G^{(1)}&=-\frac{\alpha_s}{96\pi m_\chi^3\Delta}\Big[g_1(y_t\cos\theta,m_t,m_1)+g_1(y_t\sin\theta,m_t,m_2)\nn\\
&+g_1(y_u\sin\theta,0,m_1)+g_1(y_u\cos\theta,0,m_2)+g_1(y_u,0,M_2)\Big],\\
g_G^{(2)}&=\frac{\alpha_s}{48\pi m_\chi^3\Delta^2}\Big[g_2(y_t\cos\theta,m_t,m_1)+g_2(y_t\sin\theta,m_t,m_2)\nn\\
&+g_2(y_u\sin\theta,0,m_1)+g_2(y_u\cos\theta,0,m_2)+g_2(y_u,0,M_2)\Big],
\end{align}
where
\begin{widetext}
\begin{align}
g_1(g_\chi,m_q,m_\phi) = &g_\chi^2\Big\{  [ 5m_\chi^4 m_q^2+m_\chi^4 m_\phi^2 -3 m_\chi^2 m_q^4+3m_\chi^2m_\phi^4 
+(m_q^2-m_\phi^2)^3 -3m_\chi^6]L -6m_\chi^4 
+2 m_\chi^2(m_q^2-m_\phi^2) + \Delta\ln\frac{m_q^2}{m_\phi^2}\Big\}, \\
g_2(g_\chi,m_q,m_\phi)=&g_\chi^2 \Big\{[ m_\chi^8(m_q^2-5m_\phi^2)  
+2 m_\chi^6( 2m_q^2m_\phi^2 -4 m_q^4 + 5 m_\phi^4 ) +10 m_\chi^4(m_q^6-m_\phi^6)-5m_\chi^2(m_q^2-m_\phi^2 )^3(m_q^2+m_\phi^2) \nn\\
&+( m_q^2-m_\phi^2)^5+m_\chi^{10} ]L  + 2m_\chi^6(m_q^2-4m_\phi^2) 
 7m_\chi^4(m_\phi^4-m_q^4) +2m_\chi^2(m_q^2-m_\phi^2)^3 + 3m_\chi^8  + \Delta^2\ln\frac{m_\phi^2}{m_q^2} \Big\}.
\end{align}
\end{widetext}
The definition of $\Delta$ and $L$ could be found in the Appendix; see Eqs.~\ref{eq:delta} and ~\ref{eq:L}.


Next, we perform a numerical calculation for $\sigma_{\chi N}^{\rm SD}$ and $\sigma_{\chi N}^{\rm SI}$. The hadronic matrix elements  are given by~\cite{Hill:2014yxa}{\color{yellow}:}
\begin{align}
&f_{Tu}^p=0.018,~~f_{Td}^p=0.030,~~
f_{Tu}^n=0.015,\nn\\
&f_{Td}^n=0.034,~~ f_{TG}=0.80.
\end{align}
Here $f_{Tq}^N$ corresponds to the contribution of the quark $q$ to the nucleon matrix elements for the nucleon $N$.
The matrix elements of the twist-2 operators are related to the second moments of the parton distribution functions (PDFs),
\begin{align}
q(2)+\bar{q}(2)=\int_0^1dx x(q(x)+\bar{q}(x)),~~G(2)=\int_0^1 dxxg(x),
\end{align}
where $q(x),~\bar{q}(x)$ and $g(x)$ are the PDFs of quark, anti-quark and gluon in nucleon $N$, respectively. Those hadronic elements could be extracted from the CT14NNLO PDFs~\cite{Dulat:2015mca}, 
\begin{align}
&[u(2)+\bar{u}(2)]_p=0.3481,~~[d(2)+\bar{d}(2)]_p=0.1902,\nn\\
&G(2)_p=G(2)_n=0.4159.
\end{align}
For the spin dependent matrix elements, we utilize the values in Ref.~\cite{Freytsis:2010ne},
\begin{align}
\Delta_p^u=0.84,~~\Delta_p^d=-0.43,~~\Delta_n^u=-0.43,~~\Delta_n^d=0.84.
\end{align}
\begin{figure*}
\centering
\includegraphics[scale=0.5]{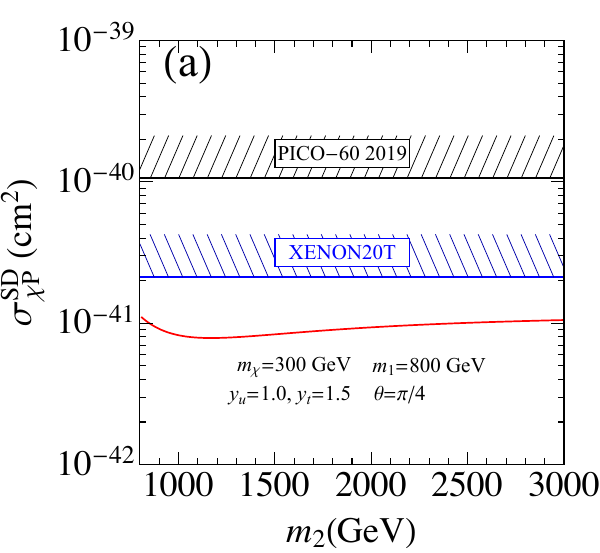}
\includegraphics[scale=0.5]{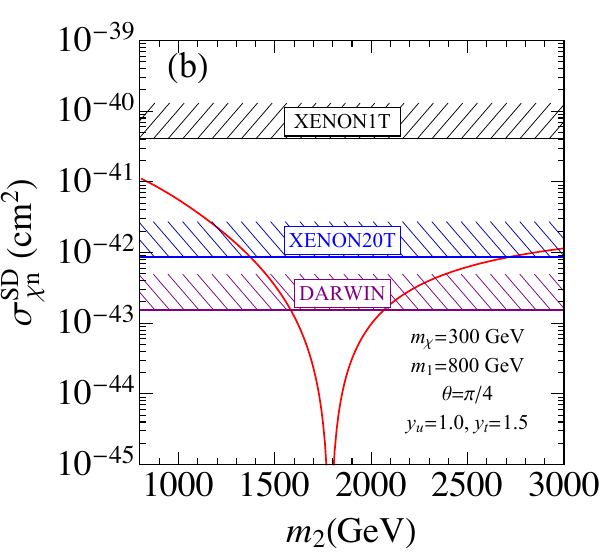}
\includegraphics[scale=0.5]{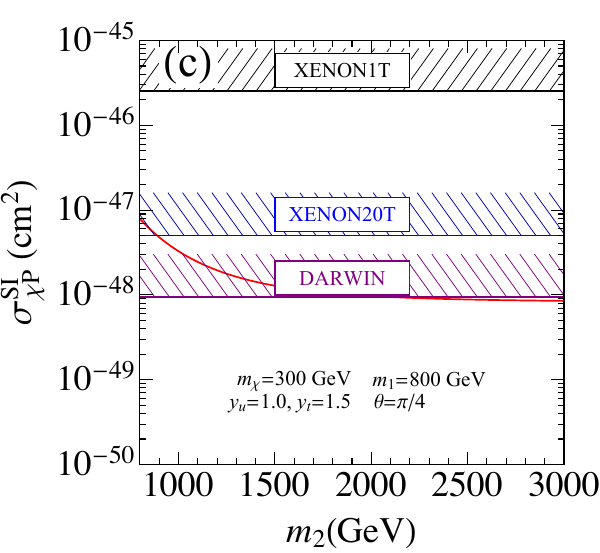}
\includegraphics[scale=0.5]{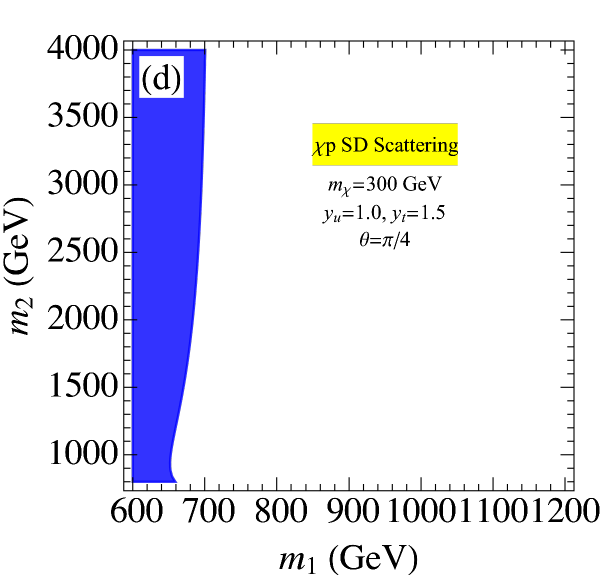}
\includegraphics[scale=0.5]{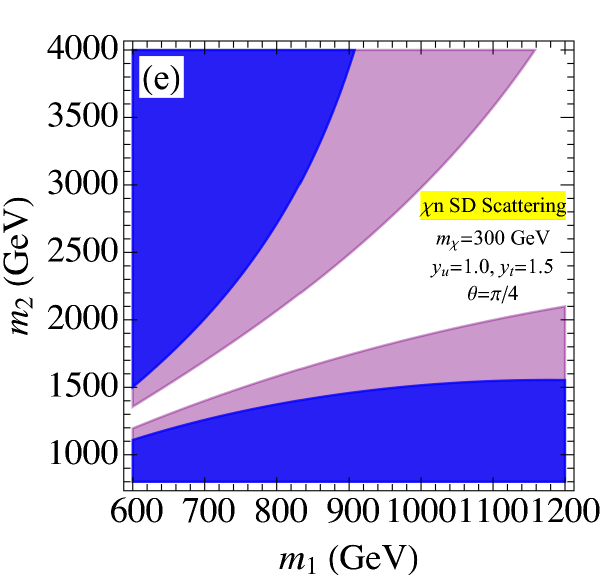}
\includegraphics[scale=0.5]{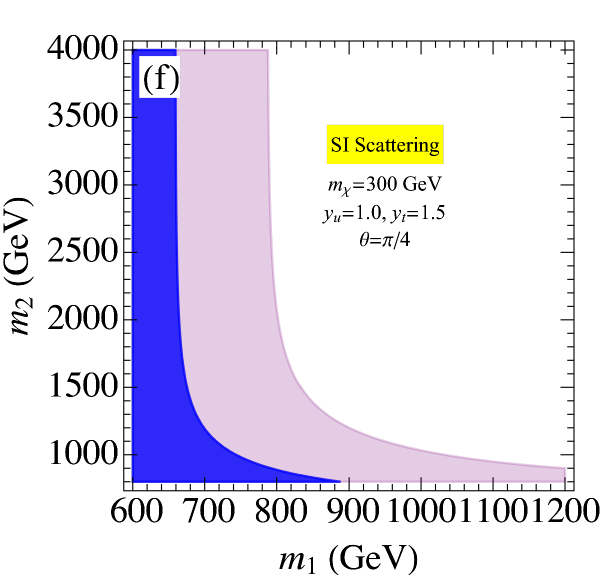}
\caption{\label{Fig:DMDD}The constraints from DM direct detection experiments on the parameter $m_2$ (a,b,c) and $m_1$-$m_2$ plane (d,e,f). The black meshed region denotes the limits from the current measurements PICO-60 and XENON1T, while the the blue and purple meshed (shaded) regions are the bound from the future XENON20T and DARWIN, respectively.}
\end{figure*}
We show the DM SD and SI scattering rates in Figs.~\ref{Fig:DMDD}(a)-(c) (red lines), with the benchmark parameters $m_{\chi}=300~{\rm GeV}$, $m_1=800~{\rm GeV}$, $y_u=1,~y_t=1.5$ and $\theta=\pi/4$.  The most severe limits for the SD scattering with proton and neutron targets are from PICO-60~\cite{Amole:2019fdf}  and XENON1T~\cite{Aprile:2018dbl}, respectively, and the upper limits are roughly $\mathcal{O}(10^{-40}{\rm cm}^2)$ (black meshed regions). The SI scattering measurement from  XENON1T could give a stronger bound for the cross section~\cite{Aprile:2018dbl} (black meshed region) , i.e. $\sigma_{\chi p}^{\rm SI}\sim 10^{-45} {\rm cm}^2$.  While the expected constraints from XENON20T~\cite{Aprile:2020vtw} (blue regions) and DARWIN~\cite{Aalbers:2016jon} (purple region) could be further improved as compared to the present data and  it shows the sensitivity could be enhanced about one to two order of magnitude for both the SD and SI scattering. We observe that the predictions from our simplified model are consistent with the current measurements and also could be tested in the future experiments. In addition, we note that the cancellation between up and down quarks' contributions in DM-neutron SD scattering will obviously change the cross section compared to the DM-proton SD process; see Fig.~\ref{Fig:DMDD}(b) (red lines). In Figs.~\ref{Fig:DMDD}(d)-(f), we show the expected sensitivity regions from the future XENON20T and DARWIN experiments on the plane of $m_1$ and $m_2$.  It is clear that both the DM-neutron SD and SI scattering could give a strong constraint for our simplified model and the measurements could probe complementary regions in the parameter space.

\section{Collider Search at the LHC}
\begin{figure}
\centering
\includegraphics[scale=0.45]{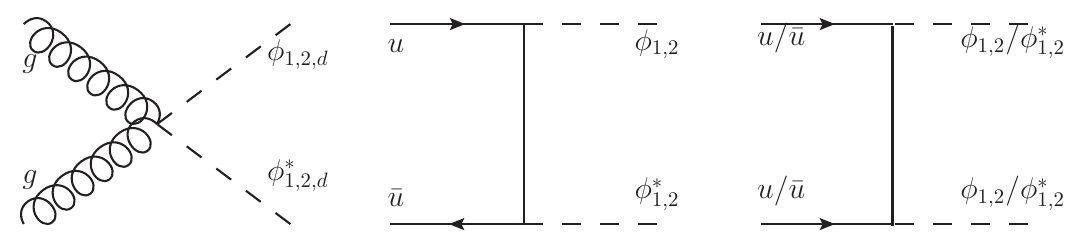}
\caption{\label{Fig:scalarproduction}Illustrative Feynman diagrams of colored scalar pair production at the LHC.}
\end{figure}

In this section we dedicate to explore the LHC limits to our simplified model. The processes of interests to us at the LHC include pair production of the colored scalars ($\phi_i$, with $i=1,2,d$) and the associated production of scalars with DM candidate $\chi$. It has been shown that the later process plays a subdominant role to constrain the model parameter space,  and will be neglected from here on~\cite{Mohan:2019zrk}. The pair production of scalars could be generated through the strong force and the Yukawa interactions; see Fig.~\ref{Fig:scalarproduction}.  We should note that those colored scalars have the same performance as the squarks in the supersymmetric models at the LHC, as a result, we can use the limits from squark searches at the LHC to estimate the bounds of our simplified model~\cite{Sirunyan:2019ctn}.

The colored scalar  $\phi_{1,2,d}$ could decay to light quark/top quark plus DM candidate $\chi$ and both of them have been  searched by ATLAS and CMS collaborations~\cite{Aad:2020aze,Aad:2020sgw,Sirunyan:2019ctn,Sirunyan:2020tyy}. It shows that the bounds from the top quark decay mode are much weaker than the light quark case in the most parameter space~\cite{Sirunyan:2019ctn}. 
Therefore, we will only consider the light quark decay mode in this work, i.e. $\phi_{1,2}\to u\chi$ and $\phi_d\to d\chi$.

We generate the parton level events at 13 TeV LHC using MadGraph5~\cite{Alwall:2014hca} and pass them to PYTHIA~\cite{Sjostrand:2014zea} for showering and hadroniation,
and simulate the collider smearing effects utilizing Delphes~\cite{deFavereau:2013fsa}. Then, following the analysis strategy of CMS collaboration~\cite{Sirunyan:2019ctn}, we require  the following kinematic cuts on the events:
\begin{align}
&n_j=2\mbox{-}3,~n_b=0,~ \mid\eta^j\mid<2.4,~ p_{\mathrm{T}}^j>30~\mbox{GeV},\nn\\
&H_{\mathrm{T}}>p_{\mathrm{T}}^{\mbox{\tiny miss}}>300\mbox{GeV},~\Delta\phi(\vec{p}_{\mathrm{T}}^{\mbox{\tiny~ miss}},j_1)>0.5\,,\nn\\
&\Delta\phi(\vec{p}_{\mathrm{T}}^{\mbox{\tiny~ miss}},j_2)>0.5\,, \Delta\phi(\vec{p}_{\mathrm{T}}^{\mbox{\tiny~ miss}},j_3)>0.3,
\end{align}
where $n_{j(b)}$ is the light (bottom) jet number, $H_{\mathrm{T}}$ is the scalar $p_{\mathrm{T}}$ sum of jets with $\mid\eta^j\mid<2.4\,$ and $p_{\mathrm{T}}^{\mbox{\tiny miss}}$ is the missing transverse momentum. The CMS search is performed in a four-dimensional region defined by exclusive intervals in $n_j$, $n_b$, $H_T$ and $p_{\mathrm{T}}^{\mbox{\tiny miss}}$. Based on the signal features of our model, we fix $n_j$ and $n_b$ in the analysis, while including the 10 kinematic intervals of  $H_T$ and $p_{\mathrm{T}}^{\mbox{\tiny miss}}$ which are defined in Ref.~\cite{Sirunyan:2019ctn}. The significance for each bin $q_{\mbox{\tiny EL}}$ could be evaluated by the likelihood method and it shows
\begin{align}
q_{\mbox{\tiny EL}}=\sqrt{-2\left[N_{\mbox{\tiny Exp}}\ln \frac{N_s+N_b}{N_b} -N_s\right]}\,, 
\end{align}
where $N_{\mbox{\tiny Exp}}\,$, $N_s$ and $N_b$ denote the number of observed, signal and SM background events, respectively. 
The $N_{\mbox{\tiny Exp}}$ and $N_b$ for each bin can be found in Ref.~\cite{Sirunyan:2019ctn} and the systematic uncertainties for the background will be ignored in this analysis. 
The total significance for the 10 bins could be obtained by square root sum of each bin. 

\begin{figure}
\centering
\includegraphics[scale=0.284]{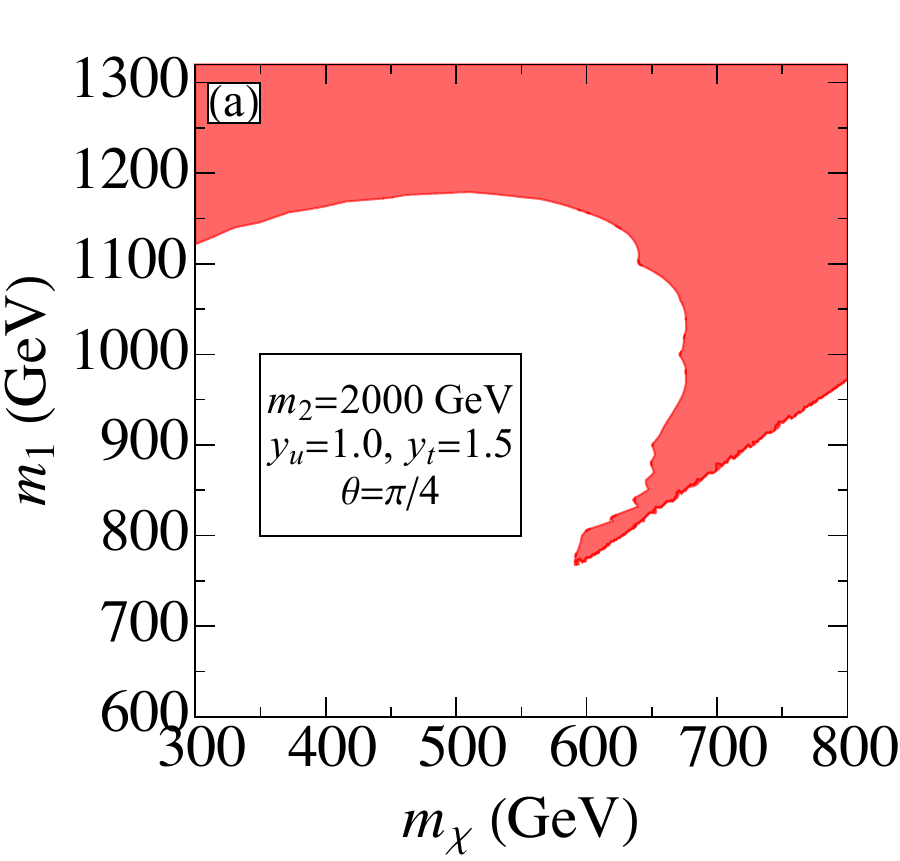}
\includegraphics[scale=0.23]{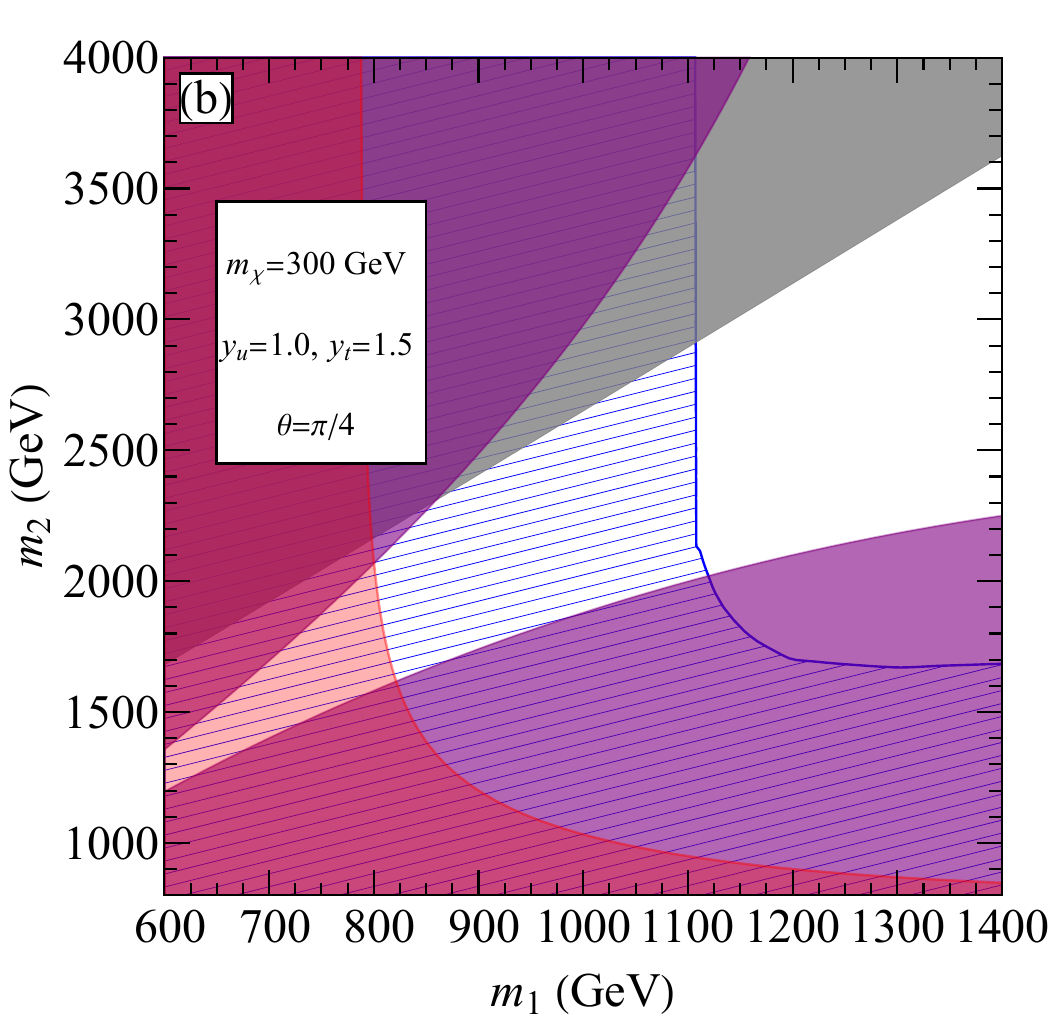}
\caption{\label{Fig:lhc}
(a) Allowed parameter space under the constraint from the 13 TeV LHC with an integrated luminosity of 137 fb$^{-1}$ at $2\sigma$  level on the plane of $m_\chi$ and $m_1$. (b) Combined constraints to the parameter space on the plane of $m_1$ and $m_2$. The blue meshed region was excluded by the 13 TeV LHC at $2\sigma$  level with an integrated luminosity of 137 fb$^{-1}$. The colored region can be probed from $tqh$ search at the FCC-hh (gray region), spin-dependent (purple region) and spin-independent (red region) DM direct detection search at the projected DARWIN experiment, respectively.
}
\end{figure}

Figure~\ref{Fig:lhc}(a) shows the allowed parameter space on the plane of $m_\chi$ and $m_1$ with benchmark parameters $m_2=2000~{\rm{GeV}}, y_u=1.0, y_t=1.5$ and $\theta=\pi/4$ at the 95\% C.L. with the integrated luminosity of 137 fb$^{-1}$ based on above analysis. 
It shows that the same-sign electric charged scalar pair production through Yukawa interaction could play a crucial role in some parameter space due to the enhancement of the up quark PDF~\cite{Acaroglu:2021qae}.
We displays the combined constraints from the DM searches at the LHC (blue meshed region), $t\to qh$ measurement from FCC-hh (gray region) and projected DM SD (purple region) and SI (red region) scattering from DARWIN on the plane of $m_1$ and $m_2$ at $2\sigma$ level in Fig.~\ref{Fig:lhc}(b).  The meshed region was excluded by the 13 TeV LHC data, while the colored regions could be probed by the future experiments.  It is clearly that the direct detection experiments, DM LHC searches and top quark FCNC measurement could probe complementary and same regions in parameter space.

Before closing this section, we would like to scan the 5-dim parameter space $\{m_{1,2,\chi}, y_{u,t}\}$ to estimate the allowed parameter space within all constraints and the results are shown in Fig.~\ref{Fig:scan}. In order to enhance the top quark FCNC effects, we choose the maximal mixing angle of $\theta=\pi/4$ in this study. We consider the following parameter space with assumption $m_\chi<m_{1,2,d}$,
\begin{align}
y_u &\in [0.4, 2.0],&y_t &\in [0.4, 2.0], &\nonumber \\
m_1 &\in [800, 2000] {~\rm{GeV}},&m_2 &\in [1000, 4000] {~\rm{GeV}},\nn\\
m_\chi &\in [300, 1200] {~\rm{GeV}}. \nonumber
\end{align}
We project the points from 5-dim space to the plane of $(m_1,m_2)$ and $(m_\chi,m_1)$. 
All points denote the parameters which are consistent with the current LHC limit and dark matter relic abundance measurement from Planck at 2$\sigma$ level. The red points describe the parameters which can be tested via $t\to qh$ at the future collider FCC-HH and the blue points imply the parameters can be crosschecked via the $tqh$ search and DM direct detection experiments from DARWIN, including both the spin-independent and spin-dependent scatterings. The black line in the plane $(m_1,m_2)$ denotes the unitary bound. It shows that both the top quark FCNC signal and DM direct detection experiments play an important role to constrain the parameter space of our simplified model.

\begin{figure}
\centering
\includegraphics[scale=0.2]{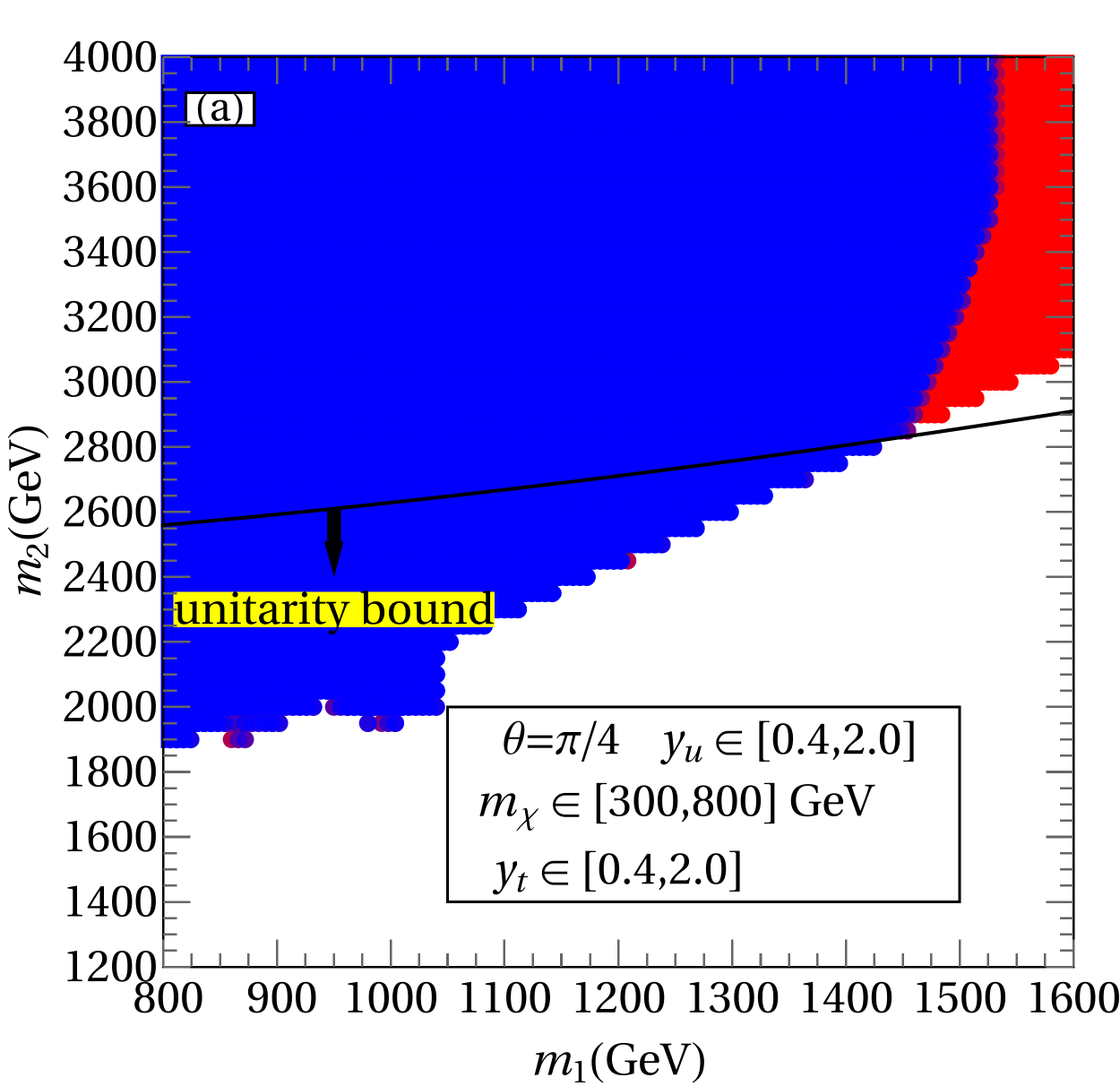}
\includegraphics[scale=0.2]{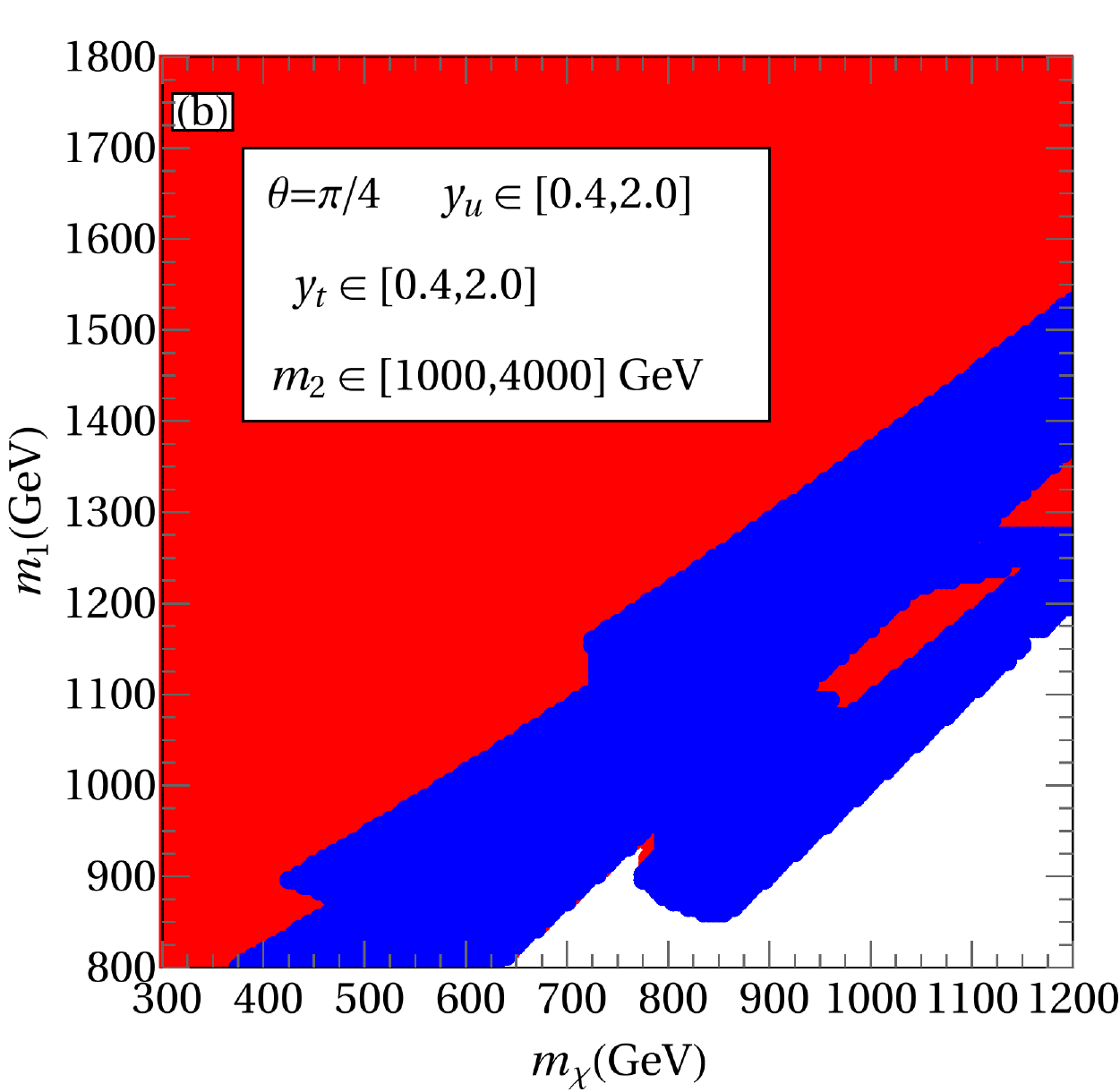}
\caption{\label{Fig:scan}
Parameters scanning in the 5-dim parameter space $\{m_1, m_2, m_\chi, y_u, y_t\}$. To maximize the top quark FCNC effects, we choose the maximal mixing angle of $\theta=\pi/4$.
The points in the figures denote the parameters which are consistent with the current LHC DM limit and DM relic abundance measurement from Planck at $2\sigma$ level. The red points can be probed via top quark FCNC coupling $tqh$ measurement at the FCC-HH, while the blue points can be probed by both the $tqh$ at the FCC-HH and DM direct detection experiment DARWIN (both spin-dependent and spin independent).} \label{Fig:scan}
\end{figure}

\section{Conclusion}
Due to the large hierarchy between the top quark FCNC anomalous couplings and the SM top quark interactions,  it is well motivated to study the loop induced top quark FCNC processes at the LHC, which could avoid the possible unduly small NP couplings in the model. In this work, we consider one class of the simplified model which  could only generate the top quark FCNC couplings in the loop level. Such scenario predicts a Majorana dark matter candidate and could be tested through the scattering between the dark matter and the nuclei, and also the data from LHC.

In our simplified model, the top quark could decay to $q+Z/\gamma/g/h$ and the most promising decay channel at the HL-LHC and future hadron colliders would be $t\to qh$. Its branching ratio is around $\mathcal{O}(10^{-5})$ in some parameter space. Comparing to the expected limits from the HL-LHC, the unitarity band from $\phi\phi\to W_LW_L$ scattering would give a stronger constraint for the model parameters.

To fully explore the discovery potential of our simplified model, we carried out a comprehensive comparsion of the dark matter relic density, direct detection experiments and LHC searches. Due to the Majorana nature of the dark matter in our model, we note that the SI  scattering with nuclei at leading order will vanish, while it could be generated at the next-to-leading order.  Although the SI cross section from higher order will be suppressed compared to the tree level SD case, the SI scattering also provides a comparable limits to the SD production.
Our analysis shows that  the simplified  model could easily satisfy the relic density limits, while the direct detection experiments from the scattering between dark matter and nuclei and LHC searches provide the severe bounds.  In particular, the constraints from scattering between dark matter and neutron. When comparing the direct detection experiments and LHC searches, we note that they could probe complementary and same regions in parameter space, and are both necessary to fully constrain our simplified model.

\begin{acknowledgments}
We thank Qing-Hong Cao, Zhaofeng Kang, Jue Zhang and Ya Zhang for helpful discussions.
The work of Y. Liu is supported in part by the National Science Foundation of China under Grant Nos. 11805013, 12075257 and the Fundamental Research Funds for the Central Universities under Grant No. 2018NTST09, BY is supported by the U.S. Department of Energy, Office of Science, Office of Nuclear Physics, under Contract DE-AC52- 06NA25396, [under an Early Career Research Award (C. Lee),] and through the LANL/LDRD Program, RZ is supported in part by the National Science Foundation of China under Grant Nos. 12075257 and the funding from the Institute of High Energy Physics, Chinese Academy of Sciences (Y6515580U1) and the funding from Chinese Academy of Sciences (Y8291120K2).
\end{acknowledgments}

\appendix 
\section{Feynman Integral}
In the appendix we represent the details of the Feynman integrals in Eq.~\ref{eq:fg},
\begin{align}
 \int \frac{d^4q}{ i\pi^2}\frac{(\slashed{p}+\slashed{q}) }{(q^2-m_\phi^2)[ (p+q)^2-m_q^2]^4},
\end{align}
and 
\begin{align}
 \int \frac{d^4q}{ i\pi^2}\frac{(\slashed{p}+\slashed{q}) }{(q^2-m_\phi^2)^4[ (p+q)^2-m_q^2]}.
\end{align}
Note we can substitute $\slashed{p}$ with $m_\chi$ in terms of Dirac equation of dark matter field with momentum $p$. Utilizing the Feynman parametrization method, we obtain \begin{align}
& \int \frac{d^4q}{ i\pi^2}\frac{(\slashed{p}+\slashed{q}) }{(q^2-m_\phi^2)[ (p+q)^2-m_q^2]^4} \nonumber \\
 =& \frac{6 m_q^2m_\phi^2(m_\phi^2+m_q^2-m_\chi^2) L +\Delta-12m_q^2m_\phi^2}{6m_q^2 \Delta^2},
\end{align}
and 
\begin{align}
& \int \frac{d^4q}{ i\pi^2}\frac{(\slashed{p}+\slashed{q}) }{(q^2-m_\phi^2)^4[ (p+q)^2-m_q^2]}\nonumber \\
=&\frac{(\Delta+6m_\phi^2m_q^2)(m_\phi^2+m_q^2-m_\chi^2)-12m_q^4m_\phi^4 L}{6m_\phi^4 \Delta^2},
\end{align}
where the symbols $\Delta$ and $L$ are defined as
\begin{align}
\Delta = 4 m_\phi^2m_\chi^2 -(m_q^2-m_\chi^2-m_\phi^2)^2,
\label{eq:delta}
\end{align}
and 
\begin{align}
L= 
\begin{cases}
\frac{2}{\sqrt{|\Delta|}}\tanh^{-1}\frac{\sqrt{|\Delta|}}{m_q^2+m_\phi^2-m_\chi^2},~~\Delta<0 \\
\frac{4m_\chi^2}{m_q^4+m_\phi^4-m_\chi^4-2m_q^2m_\phi^2},~~~~~~~~~\Delta=0 \\
\frac{2}{\sqrt{\Delta}}\tan^{-1}\frac{\sqrt{\Delta}}{m_q^2+m_\phi^2-m_\chi^2}.~~~~~~\Delta>0
\end{cases}.
\label{eq:L}
\end{align}

\bibliographystyle{apsrev}
\bibliography{reference}

\end{document}